%% file: main.tex
\documentclass[fleqn,usenatbib]{mnras}

\usepackage{newtxtext,newtxmath}

\usepackage{cprotect}
\usepackage[T1]{fontenc}

\DeclareRobustCommand{\VAN}[3]{#2}
\let\VANthebibliography\thebibliography
\def\thebibliography{\DeclareRobustCommand{\VAN}[3]{##3}\VANthebibliography}

\usepackage{graphicx}	
\usepackage{amsmath}	
\usepackage{booktabs}
\usepackage{enumitem}

\title[The SMF of quiescent galaxies in $2<z<2.5$ protoclusters]{The stellar mass function of quiescent galaxies in $2<z<2.5$ protoclusters}

\author[Adit H. Edward]{Adit H. Edward$^{1,2}$, 
Michael L. Balogh$^{1,2}$, Yannick M. Bah{\'e}$^{3,4}$,  M. C. Cooper$^{5}$, Nina A. Hatch$^{6}$, 
\newauthor Justin Marchioni$^{1,2}$, Adam Muzzin$^{4,7}$, Allison Noble$^{8}$, Gregory H. Rudnick$^{9}$, Benedetta Vulcani$^{10}$, 
\newauthor Gillian Wilson$^{11}$, Gabriella De Lucia$^{12}$, Ricardo Demarco$^{13}$, Ben Forrest$^{14}$, 
Michaela Hirschmann$^{15}$, \newauthor
Gianluca Castignani$^{16,17}$
Pierluigi Cerulo$^{18}$, 
Rose A. Finn$^{19}$, Guillaume Hewitt$^{1,2}$, Pascale Jablonka$^{20,21}$, \newauthor
Tadayuki Kodama$^{22}$,
Sophie Maurogordato$^{23}$, 
Julie Nantais$^{24}$, Lizhi Xie$^{25}$
\\
$^{1}$Department of Physics and Astronomy, University of Waterloo, Waterloo, Ontario N2L 3G1, Canada \\
$^{2}$Waterloo Centre for Astrophysics, University of Waterloo, Waterloo, Ontario, N2L3G1, Canada \\
$^{3}$Institute of Physics, Laboratory of Astrophysics, Ecole Polytechnique F{\'e}d{\'e}rale de Lausanne (EPFL), Observatoire de Sauverny, 1290 Versoix, Switzerland \\
$^{4}$Leiden Observatory, Leiden University, P.O. Box 9513, 2300 RA Leiden, The Netherlands \\
$^{5}$Department of Physics and Astronomy, University of California, Irvine, 4129 Frederick Reines Hall, Irvine, CA 92697, USA \\
$^{6}$School of Physics and Astronomy, University of Nottingham, Nottingham NG7 2RD, UK \\
$^{7}$Department of Physics and Astronomy, York University, 4700, Keele Street, Toronto, ON, MJ3 1P3, Canada \\
$^{8}$School of Earth and Space Exploration, Arizona State University, Tempe, AZ 85287, USA \\
$^{9}$Department of Physics and Astronomy, University of Kansas, Lawrence, KS 66045, USA \\
$^{10}$INAF - Osservatorio Astronomico di Padova, Vicolo Osservatorio 5, I-35122 Padova, Italy \\
$^{11}$Department of Physics, University of California Merced
5200 Lake Road, Merced, CA 95343, USA\\
$^{12}$INAF - Astronomical Observatory of Trieste, via G.B. Tiepolo 11, I-34143 Trieste, Italy \\
$^{13}$Departamento de Astronom\'ia, Facultad de Ciencias F\'isicas y Matem\'aticas, Universidad de Concepci\'on, Casila 160-C, Concepci\'on, Chile \\
$^{14}$Department of Physics and Astronomy, University of California Davis, One Shields Avenue, Davis, CA, 95616, USA \\
$^{15}$Institute for Physics, Laboratory for Galaxy Evolution and Spectral modelling, Ecole Polytechnique Federale de Lausanne, Observatoire de Sauverny, Chemin Pegasi 51, 1290 Versoix, Switzerland\\
$^{16}$Dipartimento di Fisica e Astronomia ”Augusto Righi”, Alma Mater Studiorum Università di Bologna, Via Gobetti 93/2, I-40129 Bologna, Italy \\
$^{17}$INAF - Osservatorio di Astrofisica e Scienza dello Spazio di Bologna, via Gobetti 93/3, I-40129, Bologna, Italy \\
$^{18}$Departamento de Ingenier\'ia Inform\'atica y Ciencias de la Computaci\'ion, Facultad de Ingenier\'ia, Universidad de Concepci\'on,   Casila 160-C, Concepci\'on, Chile \\
$^{19}$Department of Physics and Astronomy, Siena College, 515 Loudon Rd, Loudonville, NY 12211, USA \\
$^{20}$Laboratoire d’astrophysique, École Polytechnique Fédérale de Lausanne (EPFL), 1290 Sauverny, Switzerland \\
$^{21}$GEPI, Observatoire de Paris, Université PSL, CNRS, Place Jules Janssen, F-92190 Meudon, France \\
$^{22}$Astronomical Institute, Tohoku University, 6-3, Aramaki, Aoba, Sendai, Miyagi 980-8578, Japan \\
$^{23}$Observatoire de la Côte d'Azur, CNRS, Laboratoire Lagrange, Bd de l'Observatoire, Université Côte d'Azur, CS 34229, 06304, Nice Cedex 4, France \\
$^{24}$Instituto de Astrofísica, Departamento de Ciencias Físicas, Universidad Andres Bello, Fernandez Concha 700, Las Condes 7591538, Santiago, Región Metropolitana, Chile \\
$^{25}$Tianjin Astrophysics Center, Tianjin Normal University, Binshuixidao 393, 300387, Tianjin, People’s Republic of China \\
}

\date{Accepted XXX. Received YYY; in original form ZZZ}

\pubyear{2023}

\begin{document}
\label{firstpage}
\pagerange{\pageref{firstpage}--\pageref{lastpage}}
\maketitle

\begin{abstract}
We present an analysis of the galaxy stellar mass function (SMF) of 14 known protoclusters between $2.0 < z < 2.5$ in the COSMOS field, down to a mass limit of $10^{9.5}~\text{M}_{\odot}$.  We use existing photometric redshifts with a statistical background subtraction, and consider star-forming and quiescent galaxies identified from $(NUV-r)$ and $(r-J)$ colours separately.  Our fiducial sample includes galaxies within $1$~Mpc of the cluster centres. 
The shape of the protocluster SMF of star-forming galaxies is indistinguishable from that of the general field at this redshift. Quiescent galaxies, however, show a flatter SMF than in the field, with an upturn at low mass, though this is only significant at $\sim2\sigma$. 
There is no strong evidence for a dominant population of quiescent galaxies at any mass, with a fraction $<15$\% at $1\sigma$ confidence for galaxies with $\log{M_{\ast}/M_\odot}<10.5$.   
We compare our results with a sample of galaxy groups at $1<z<1.5$,
and demonstrate that a significant amount of environmental quenching must take place between these epochs, increasing the relative abundance of high-mass ($\rm M_{\ast} > 10^{10.5} M_{\odot}$) quiescent galaxies by a factor $\gtrsim 2$. However, we find that at lower masses ($\rm M_{\ast} < 10^{10.5} M_{\odot}$), no additional environmental quenching is required.
\end{abstract}

\begin{keywords}
Galaxies: evolution -- Galaxies: clusters:general 
\end{keywords}



\defcitealias{Weaver_2022}{W22}
\defcitealias{Weaver_SMF}{W23}

\section{Introduction}
Observations of the galaxy stellar mass function (SMF), primarily from photometric redshift surveys, have demonstrated that most of the stellar mass in the Universe forms by $z\sim 2$ \citep[e.g.][]{D+03, GR+03, GR+06, Ilbert_2013, Muzzin_2013,  Davidzon_2017, Leja_2020, McLeod_2021, Santini_2022, Weaver_SMF, Taylor_2023}.
These studies have shown that the shape of the SMF for star-forming galaxies alone evolves only weakly with redshift below $z \sim 2$, and therefore that subsequent growth via star formation must cease for a significant number of galaxies.  This process, known as quenching, leads to a gradual accumulation of non-star forming, passively-evolving galaxies. 
\cite{Peng_2010} showed that the evolution of SMFs over $0 < z < 2$ can be matched by an empirical model in which galaxies quench with a probability that is proportional to their star formation rate (SFR).  Other authors have shown that this can be achieved with a quenching probability that is more fundamentally related to halo mass \citep{DB06}. This empirical model is often referred to as mass quenching, and is likely driven in part by energy injection due to AGN \citep{silk1998quasars, Hopkins_2006} and supernova feedback \citep{Dekel_1986,Ceverino_2009}. 

Galaxies are also affected by their environment, and processes like ram pressure stripping \citep{1972ApJ...176....1G,Poggianti_2017}, starvation \citep{LTC,Balogh2000} and galaxy mergers \citep{2005ApJ...620L..79S} can lead to environmental quenching and an excess fraction of passive galaxies in high-density regions \citep[e.g.][]{Blanton_2005, Wetzel_2012}, such as galaxy clusters \citep{Lewis_2002,Gomez_2003} and galaxy groups \citep{SM+11}. 

\citet{Baldry_2006} and \citet{Peng_2010} showed that the fraction of quenched galaxies at $z=0$ depends separably on mass and environment. The simplest interpretation is that the effectiveness of environmental quenching is independent of galaxy mass.
However, observations have shown that this separability does not hold at higher redshifts \citep[e.g.][]{K+17,PC+19}.  For example, an analysis of clusters at $0.8<z<1.5$ in the GCLASS (Gemini CLuster Astrophysics Spectroscopic Survey) and GOGREEN (Gemini Observations of Galaxies in Rich Early Environments) surveys \citep{Balogh_2020} shows that the excess quenched fraction in clusters relative to the field is strongly mass dependent \citep{Balogh_2016, van_der_Burg_2020}.  In particular, for massive galaxies only, the excess of quenched galaxies relative to the field is as high as it is in the local Universe.  The stellar populations in these galaxies are also very old \citep{Webb_2020}, indicating that they likely ceased forming stars long before they were part of a rich cluster.  This is consistent with earlier work by \citet{Thomas_2005}, who show that most star formation in early-type galaxies located in high-density environments is expected to have happened between $3 < z < 5$.  This may partly be attributed to a ``preprocessing'' that occurs in groups and filaments long before galaxies are accreted into massive clusters \citep[e.g.][]{Reeves,Werner}. 
Alternatively, or in addition, there may be a "primordial" population of massive quiescent galaxies that were formed during the very earliest stages of cluster assembly \citep[see also][]{Poggianti_2006}.  For quiescent galaxies with lower stellar mass, $\lesssim 10^{10.5}~{\rm M}_\odot$, there is strong evidence that their star formation ceased much later, upon first infall into a massive cluster \citep{Muzzin2014,McNab_2021}, leading to a more gradual build up of quiescent galaxies in clusters \citep[e.g.][]{GB08}.  Alternatively, \citet{Baxter_2022,Baxter_2023} showed that an accretion-based quenching model could work at all masses if the quenching timescale is dependent on mass, such that massive galaxies quench more quickly and earlier than less massive galaxies.

By definition, primordial quenching would have occurred within protoclusters -- the overdense, pre-virialized volumes at $z\gtrsim 2$ that will eventually collapse and form massive clusters. These volumes are very large, and only modestly overdense \citep{Muldrew_2015, Chiang_2017}. Direct observation of the galaxy population in these regions is required to decouple the primordial quiescent population from later accretion-driven quenching.   This is challenging, as it requires a survey of galaxies over a wide area that is unbiased (e.g. with respect to SFR and dust content) down to a sufficiently low stellar mass in order to study the regime at which accretion driven quenching is dominant. The most accurate way to identify protocluster members is exploiting a highly complete, deep spectroscopic survey above $z > 2$, which does not yet exist. Though there have been spectroscopic observations of protoclusters above this redshift \citep[e.g.,][]{Yuan_2014, Lee_2016, Wang_2016, Diener_2015, Darvish_2020, McConachie_2022, Ito_2023}, these are insufficient in completeness, spatial extent and depth. The alternative is to use photometric redshifts.  The larger uncertainties associated with these redshifts, however, mean large samples are required so that the signal from these modest overdensities can be extracted in the presence of a dominant background.

For this reason we use the data from The Cosmic Evolution Survey \citep[COSMOS,][]{2007ApJS..172....1S}, the survey with the best photometric redshifts over a cosmologically significant area. More specifically, we take advantage of the deep ($\sim$ 26 AB) multi-band photometry from the COSMOS2020 catalogue \citep[hereafter \citetalias{Weaver_2022}]{Weaver_2022}, covering $\sim$ 2 deg$^2$.  
In this paper, we analyze the SMFs of quiescent and star-forming galaxies within 14 previously identified protoclusters in this field, selected from the catalogue of \citet{Ata_2022} to be at $2.0 < z < 2.5$.  In constructing the SMFs we largely follow the methodology described in \citep[hereafter \citetalias{Weaver_SMF}]{Weaver_SMF}.

This paper is structured as follows. In Section \ref{Data}, we discuss the galaxy sample selection and stellar mass completeness, as well as how we select protocluster members given the photometric redshift precision.   Our methodology for constructing the SMFs is presented in Section \ref{Analysis}, and the results are described in Section \ref{Results}.  In Section~\ref{Discussion} we discuss the implications of our findings, including a comparison with plausibly descendent $1<z<1.5$ group SMFs from \citet{Reeves}.

All magnitudes are presented in the AB magnitude system \citep{1974ApJS...27...21O}. We used the `vanilla' $\Lambda$CDM cosmology model ($\Omega_m = 0.3, \Omega_{\Lambda} = 0.7, \text{H}_{0} = 70~\text{km}~\text{s}^{-1}~\text{Mpc}^{-1}$). Stellar mass estimates are taken from COSMOS2020, which assumes a \citet{Chabrier} inital mass function. We present uncertainties at the 1$\sigma$ level unless otherwise specified.

\section{Data} \label{Data}

\subsection{COSMOS2020 Sample Selection} 
\label{Sample Selection}\label{NUVrJ}

\begin{figure}
\centering
\includegraphics[width=1\linewidth]{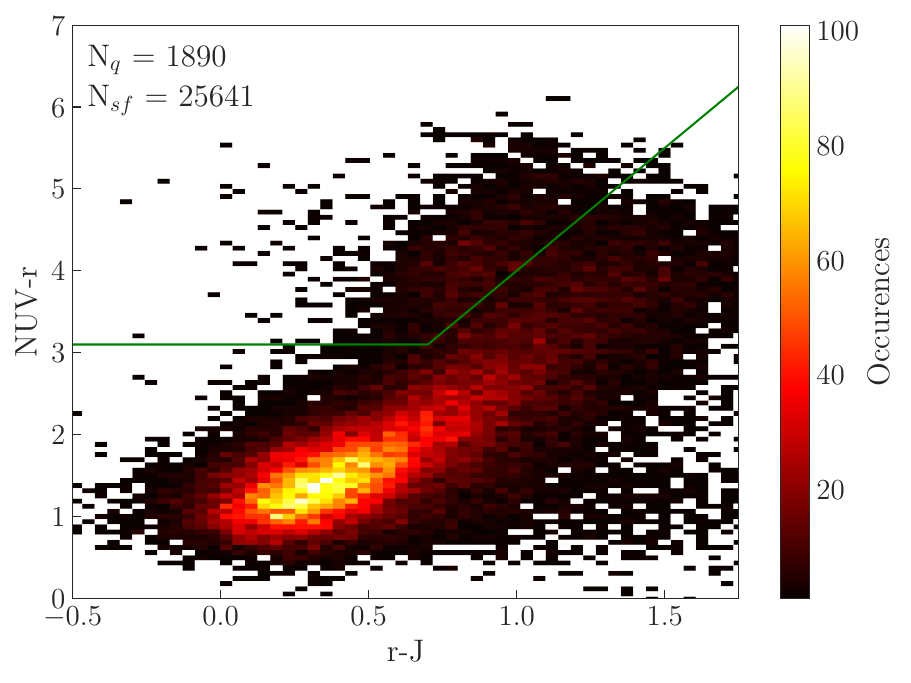}
\caption{We show the NUVrJ colour-colour diagram for all galaxies in our sample between $1.8 < z < 2.7$ above our mass limit (see Sec. \ref{mass completeness}). The solid line shows the division between the quiescent and star-forming galaxies \citep{Ilbert_2013}, with the quiescent population being above this line and the star forming one below it. We use this definition throughout this analysis. N$_{q}$ and N$_{sf}$ are the number of quiescent and star-forming galaxies identified by this criterion, respectively.}
\label{fig:NUVrJ_all}
\end{figure}

Our data are taken from the COSMOS2020 catalogue \citepalias{Weaver_2022}, based on detections in a combined $izYJHK_{s}$ image. We restrict our analysis to data within the UltraVISTA survey footprint \citep{McCracken_2012} that are not in the bright star HSC-SSP PDR2 mask nor in the bright star Suprime-Cam mask. This region corresponds to $\sim$ 1.278 $\text{deg}^2$ and is flagged in the catalogue as \verb|FLAG_COMBINED == 0|. We also limit our sources to ones with photometry measured by \verb|THE FARMER| algorithm. 
\verb|THE FARMER|, henceforth simply \verb|Farmer|, is a software package that uses \verb|The Tractor| \citep{2016ascl.soft04008L} to model and create a full multi-wavelength catalog. Specifically, we take the photometric redshifts, mass and rest-frame magnitude measurements from the \verb|LePhare| \citep{2011ascl.soft08009A} in combination with \verb|Farmer|. This is because this combination has been noted to have the best photo-$z$ performance \citepalias{Weaver_2022, Weaver_SMF}.

This region selection leaves us with a subset of the catalog with 746976 entries. When we restrict this sample to galaxies between $1.8 < z < 2.7$, we are left with 105664 entries. This choice of redshift range is informed by the precision of COSMOS2020 photometric redshifts (see Sec. \ref{photoz}) around our protocluster sample ($2 < z < 2.5$; see Sec. \ref{membership}). 
We then select all objects that are above the 5$\sigma$ IRAC channel 1 magnitude limit of 26 to ensure reliable stellar mass measurements. This magnitude cut removes 23726 objects, leaving 81938 galaxies. While this is a large cut, \citetalias{Weaver_SMF} note that $\sim 93\%$ of these sources are below our optimistic mass limit and thus will be excluded anyway (see Sec. \ref{mass completeness}). 
To remove objects with poor photometric redshifts, we restrict our analysis to ``good" fits (\verb|lp_chi2_best < 5|), removing another 779 galaxies (0.95\%). We also require \verb|lp_zPDF_u68| and \verb|lp_zPDF_l68|, the upper and lower 68 percentile confidence levels of the photometric redshift respectively, to differ by $<1.0$ to ensure that our photometric redshifts are relatively accurate, further removing 1653 galaxies (2.0\%) and leaving us with 79506 galaxies in our sample.


\begin{figure*}
\centering
\includegraphics[width=.95\linewidth, trim = 150 0 225 0]{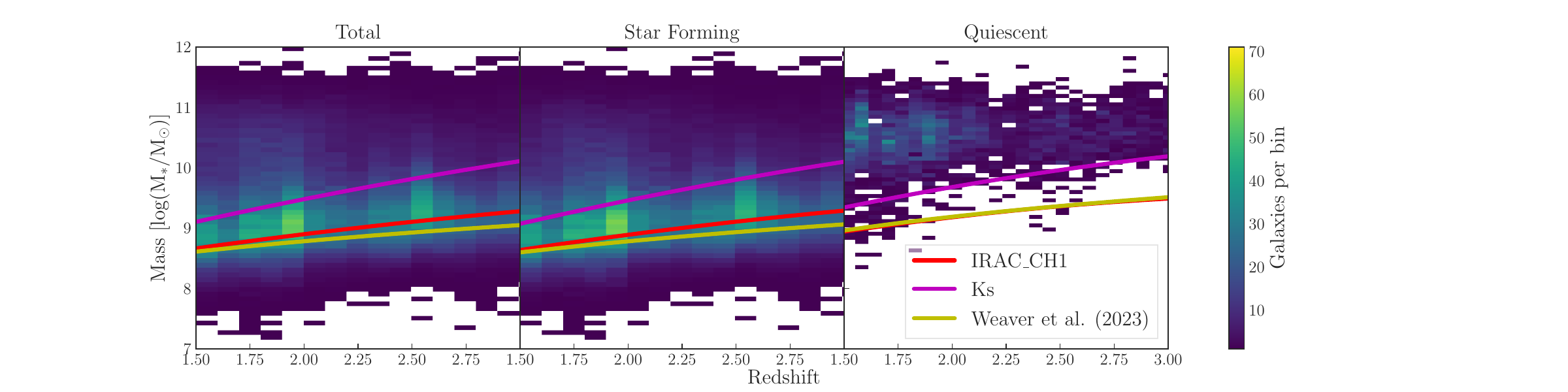}
\cprotect\caption{Stellar mass completeness as a function of redshift for our COSMOS2020 subsample in the left panel, with the sample restricted to star-forming and quiescent galaxies in the middle and right panels, respectively.  The colour indicates the number of galaxies in each bin of redshift and stellar mass. The mass completeness is determined following \citetalias{Weaver_SMF}, based on either the \verb|IRAC_CH1| magnitude limit (optimistic, shown as the red line) or the $K_s$ magnitude limit (conservative, shown as the magenta line), as described in the text.  
This is compared with the \verb|IRAC_CH1|-based completeness from \citetalias{Weaver_SMF}, shown as the yellow-green line. At the furthest redshift considered in this analysis ($z \sim 2.7$), we are complete down to $\sim 10^{10} M_{\odot}$ in our most conservative limit, and complete down to $\sim 10^{9.5} M_{\odot}$ in our optimistic limit.} 
\label{fig:mass_complete}
\end{figure*}
\begin{table*}
\begin{tabular}{lllcc}
\textbf{Population}   & \multicolumn{2}{c}{\textbf{log$_{10}$(M$_{\rm lim}$/M$_{\odot}$)} at $z=2.7$} & \multicolumn{2}{c}{\textbf{log$_{10}$(M$_{\rm lim}$/M$_{\odot}$) fit}}          \\
 & $\verb|IRAC_CH1|$ & \textbf{$K_s$} & $\verb|IRAC_CH1|$ & \textbf{$K_s$} \\
 \hline
Quiescent    & 9.5  & 10.0  & $-0.11(1+z)^2 + 1.08(1+z) + 6.93$ & $-0.11(1+z)^2 + 1.28(1+z) + 6.83$ \\
Star-Forming & 9.1  & 9.9  & $-0.07(1+z)^2 + 0.89(1+z) + 6.85$ & $-0.09(1+z)^2 + 1.27(1+z) + 6.46$ \\
Total        & 9.1  & 9.9  & $-0.06(1+z)^2 + 0.80(1+z) + 7.04$ & $-0.08(1+z)^2 + 1.19(1+z) + 6.63$
\end{tabular}
\cprotect\caption{Stellar mass limits for each population. We first show both the \verb|IRAC_CH1| -based and $K_{s}$ -based  mass limits, evaluated at $z = 2.7$ for each population. We also show each mass limit as a polynomial function of redshift. }
\label{tab:Mlim_table}
\end{table*}

Colour-colour diagnostics are effective at separating dusty star-forming galaxies from quiescent ones \citep{Arnouts_2007, Ilbert_2013}.  We use rest-frame colours provided in the \citetalias{Weaver_2022} catalogue.  In Figure~\ref{fig:NUVrJ_all} we show the (NUV - r) and (r - J) colour distribution for our sample. The use of rest NUV magnitudes in this diagnostic provides more sensitivity to age than the typical UVJ diagrams \citep{Martin_2007, Arnouts_2007}.  To split the total population into quiescent and star-forming galaxies, we use the definition of \cite{Ilbert_2013} where galaxies with rest-frame colours such that $(NUV - r) > 3(r - J) + 1$ and $(NUV - r) > 3.1$ (hereby referred to as NUVrJ selection) are considered quiescent.  This selection approximates a cut in sSFR $\lesssim$ $10^{-11}$ yr$^{-1}$ \citep{Ilbert_2013,Davidzon_2017}, and is shown as the green line in Fig. \ref{fig:NUVrJ_all}.  Note that at this redshift, the bimodality in colour distribution is still apparent, though the two populations are not completely disjoint.  We also caution that galaxies with very recently terminated star formation may still be classified as star-forming using the NUVrJ method \citep[e.g.][]{McConachie_2022}.

\subsection{Mass Completeness Limit} \label{mass completeness}
To find the mass completeness of the subset of COSMOS data used in this analysis, we take a similar approach to \citetalias{Weaver_SMF}, as originally presented in \cite{Pozzetti_2010}. \citetalias{Weaver_SMF} use the \verb|IRAC_CH1| limiting magnitude to estimate the mass completeness, as described further below.  This will overestimate the completeness, because red objects detected in  \verb|IRAC_CH1| may be missed in the detection image \citep[][\citetalias{Weaver_2022}]{Davidzon_2017}.  Indeed, a comparison with the deeper CANDELS \citep[The Cosmic Assembly Near-infrared Deep Extragalactic Legacy Survey;][]{Grogin+2011, Koekemoer+11} catalogue shows that, at the determined 95\% mass limit, only 75\% of CANDELS sources are recovered.  A more conservative choice is to use the $K_{\text{s}}$ band limit; this will underestimate the completeness because the deep Subaru/HSC photometry will allow the detection of galaxies below that limit.  We therefore take the approach of showing our results relative to both mass limits. Although \verb|IRAC_CH1| is a significantly better tracer of stellar mass for $z\gtrsim 2.5$, at the redshifts of interest here $K_{\text{s}}$ is still acceptable.

Following \citetalias{Weaver_SMF}, we bin the galaxies in redshift and identify a cutoff magnitude $m_{\rm cutoff}$ that corresponds to the 30$^{\rm th}$ percentile of magnitudes in that bin for each band.   
We then consider all galaxies with a magnitude fainter than $m_{\rm cutoff}$ and re-scale their masses so that their apparent magnitude in a given band matches the limiting magnitude:

\begin{equation}
    \text{log}_{10}\left(\frac{M_{\text{rescale}}}{M_{\odot}}\right) = \text{log}_{10}\left(\frac{M_{\ast}}{M_{\odot}}\right) + 0.4(m - m_{\rm lim}),
\end{equation}
Where $m$ is the magnitude in a given band, and $m_{\rm lim}$ is the limiting magnitude in that band (26.0 for \verb|IRAC_CH1| and 25.7 for \verb|UVISTA_Ks_MAG| \citepalias{Weaver_2022, Weaver_SMF}). We then take our limiting mass $M_{\text{\rm lim}}$ to be the 95$^{\text{th}}$ percentile of the re-scaled mass distribution in each bin and fit a polynomial to these $M_{\text{\rm lim}}$ as a function of redshift. We do this for the total, star-forming and quiescent populations.

The mass completeness of our sample compared to the one presented in \citetalias{Weaver_SMF}
is shown in Fig. \ref{fig:mass_complete}.  
Given that our analysis is restricted to protoclusters between $2 < z < 2.5$ and the furthest associated galaxies should be at roughly $z = 2.7$ (see section \ref{photoz}), we conservatively restrict our analysis to galaxies above the mass limit at this redshift. Using this, we obtain a \verb|IRAC_CH1|-based mass
completeness limit of $\log_{10}{{M}_{\rm lim}/M_\odot} = 9.1$ for the total population, 9.5 for the quiescent population, and 9.1 for the star-forming population. 

We take our aforementioned mass completeness values as our optimistic mass completeness limit. We follow the same procedure in the K$_{\text{s}}$ magnitude band (which has a limiting magnitude of 25.7) to give conservative mass completeness limits of $\log_{10}{\left({M}_{\rm lim}/M_\odot\right)}=9.9$ for both the total and star-forming populations, and $10.0$ for the quiescent population. We summarize our mass limits, both \verb|IRAC_CH1| based and K$_{\rm s}$ based in Table \ref{tab:Mlim_table}.  Our final sample for galaxies with $\text{log}_{10}\left(M_{\ast}/M_{\odot}\right)>9.5$, 
above the optimistic mass completeness limit for quiescent galaxies, consists of 27531 galaxies, of which 1890 are quiescent and 25641 are star-forming.

\begin{figure*}
\centering
\includegraphics[width=\linewidth]{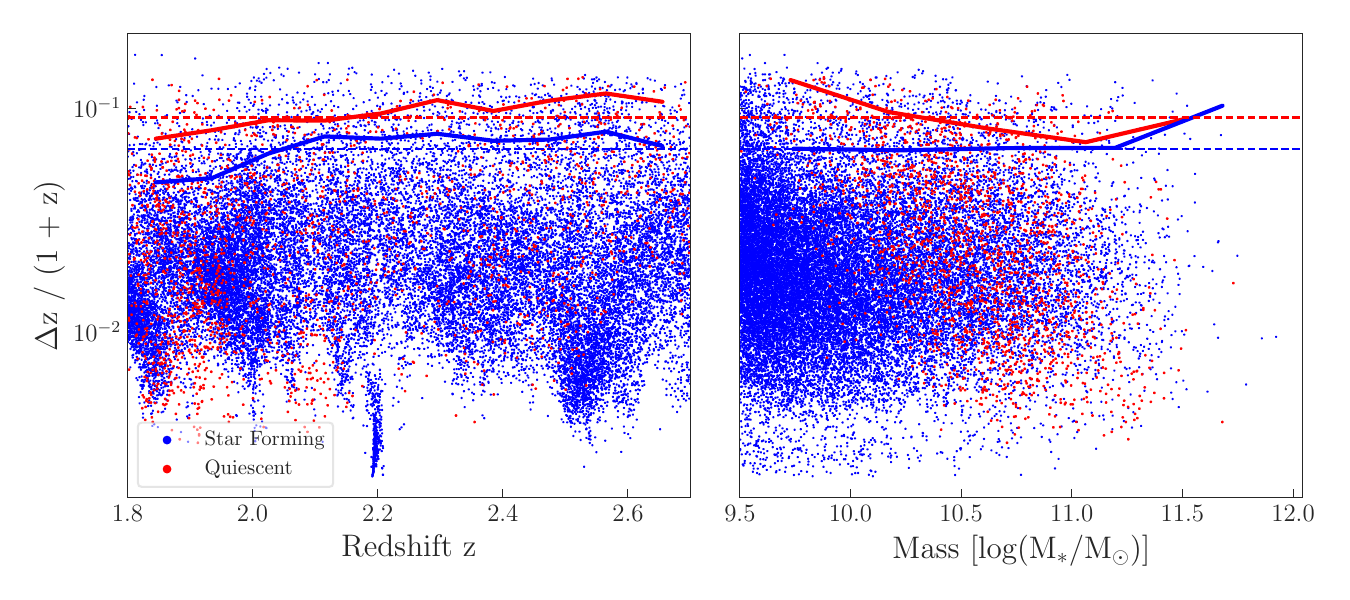}
\caption{\textit{Left:} The dependence of $\Delta z/(1+z)$ as a function of redshift and for galaxies in COSMOS2020 between $1.8 < z < 2.7$ where $\Delta z$ is defined to be the mean distance between the upper 1$\sigma$ limit and the lower one. The dashed lines are the overall upper 2$\sigma$ of the identified sample over the whole redshift range shown, while the solid lines are the upper 2$\sigma$ limit in bins of redshift. We note a slight increase in $\Delta z$ as a function of redshift.  
At the midpoint of $z = 2.25$ the upper 2$\sigma$ value is $\Delta z \approx 0.2$ for star-forming galaxies and $0.3$ for quiescent ones.
\textit{Right:} Similar, but for $\Delta z/(1+z)$ as a function of stellar mass. There is a modest decrease with increasing mass, for quiescent galaxies.
}
\label{fig:dz_vs_z_mass}
\end{figure*}

\begin{table*}
\begin{tabular}{@{}lllll@{}}
\toprule
\textbf{Protocluster Candidate} & \textbf{RA {[}deg{]}} & \textbf{Dec {[}deg{]}} & \textbf{Redshift} & \textbf{Projected $z=0$ Mass} \\ \midrule
ZFOURGE/ZFIRE                     & 150.094              & 2.251                  & 2.095   & $(1.2 \pm 0.3)\times 10^{15}~h^{-1}~{\rm M}_{\odot}$  \\
CC2.2                            & 150.197               & 2.003                  & 2.232   & $(4.2 \pm 1.9)\times 10^{14}~h^{-1}~{\rm M}_{\odot}$  \\
Hyperion 1                       & 150.093               & 2.404                  & 2.468   &          \\
Hyperion 2                       & 149.976               & 2.112                  & 2.426   &          \\
Hyperion 3                       & 149.999               & 2.253                  & 2.444   &          \\
Hyperion 4                       & 150.255               & 2.342                  & 2.469   &  $(2.5 \pm 0.5)\times 10^{15}~h^{-1}~{\rm M}_{\odot}$  \\
Hyperion 5                       & 150.229               & 2.338                  & 2.507   &          \\
Hyperion 6                       & 150.331               & 2.242                  & 2.492   &          \\
Hyperion 7                       & 149.958               & 2.218                  & 2.423   &          \\
COSTCO J100026.4                 & 150.110               & 2.161                  & 2.298   & $(4.6 \pm 2.2)\times 10^{14}~h^{-1}~{\rm M}_{\odot}$  \\
COSTCO J095924.0                 & 149.871               & 2.229                  & 2.047   & $(6.1 \pm 2.5)\times 10^{14}~h^{-1}~{\rm M}_{\odot}$  \\
COSTCO J100031.0                 & 150.129               & 2.275                  & 2.160   & $(5.3 \pm 2.6)\times 10^{14}~h^{-1}~{\rm M}_{\odot}$  \\
COSTCO J095849.4                 & 149.706               & 2.024                  & 2.391   & $(6.6 \pm 2.3)\times 10^{14}~h^{-1}~{\rm M}_{\odot}$  \\
COSTCO J095945.1                 & 149.938               & 2.091                  & 2.283   & $(4.3 \pm 2.4)\times 10^{14}~h^{-1}~{\rm M}_{\odot}$  \\ \bottomrule
\end{tabular}
\cprotect\caption{\label{tab:Protocluster_candidates} A revised version of Table 1 from \cite{Ata_2022}, providing a list of successful protocluster candidates in the COSMOS2020 field. Each candidate was identified in constrained simulations in the COSMOS field as the location of an overdensity that is likely ($> 50\%$) to evolve into a protocluster \citep{Ata_2022}. 
}
\end{table*}

\subsection{Photometric Redshifts}
\label{photoz}
We now consider the uncertainties on the photometric redshifts for the redshift range of interest ($1.8 < z < 2.7$). In the following, we define $\Delta z$ to be half the difference between the upper and lower $68\%$ confidence limits from the \verb|LePhare| code as provided by \citetalias{Weaver_2022}. These uncertainties have been shown to represent the scatter between photometric and spectroscopic redshifts well \citepalias[see ][Fig 13]{Weaver_2022}.  Although the outlier fraction becomes large ($\sim 20\%)$ at the magnitude limit of the sample, most of these outliers are high redshift ($z_{\rm spec}>3$) galaxies with $z_{\rm phot}<1$, and thus do not impact our sample selection.    However we caution that spectroscopy at $2<z<3$ is very challenging from the ground, particularly for quiescent galaxies, and samples are therefore biased.  Therefore the photometric redshift uncertainties at the magnitude limit cannot be considered to be as well characterized as for the rest of the sample.  

In Fig. \ref{fig:dz_vs_z_mass}, we show $\Delta z/(1+z)$ as a function of redshift and mass for galaxies in our sample after the selections described in Section \ref{Sample Selection}. The precision of the photometric redshifts do not depend significantly on redshift or mass in this redshift regime. For the redshift range of our sample,  $1.8 < z < 2.7$,  95$\%$ of all star-forming galaxies have $\Delta z / (1 + z) < 0.06$, which corresponds to $\Delta z \approx 0.2$ at $z = 2.25$, the midpoint redshift in this range. When considering quiescent galaxies, $95\%$ of all entries have a  $\Delta z / (1 + z) < 0.09$, which corresponds to $\Delta z \approx 0.3$ at $z = 2.25$. 

\subsection{Cluster Membership}
\label{membership}
While many $z > 2$ protocluster candidates have been identified in the literature, in general it is not possible to know for certain whether these are true protoclusters in the sense that they will evolve into massive ($>10^{14}~\text{M}_{\odot}$) virialized structures by $z = 0$. Recently, \cite{Ata_2022} analysed constrained N-body (dark matter only) simulations of the COSMOS density field, with initial fluctuations at $z=100$ chosen to evolve into the three-dimensional structure within the central square degree of the COSMOS field, as defined by extensive spectroscopic redshifts.  From fifty randomly selected realizations of these initial conditions, the simulations are evolved to $z=0$ to predict the final state of all protocluster candidates in this field.  
For the present analysis, we consider only those protoclusters that have a high probability (generally $>80\%$, with one exception) of evolving into massive clusters by $z = 0$ based on their analysis; these are listed in Table \ref{tab:Protocluster_candidates}. We start with a summary of each protocluster, though more details can be found in \cite{Ata_2022}:

\begin{description}
    \item [ZFOURGE/ZFIRE:] This system was first discovered using a near-IR imaging survey with five custom medium-bandwith filters \citep{Spitler_2012}, and was then confirmed by a spectroscopic follow up \citep{Yuan_2014}. It was measured to have a velocity dispersion of $\sigma = 552 \pm 52$ km s$^{-1}$ \citep{Yuan_2014}. In all 50 runs of the constrained simulations, this protocluster was found to evolve into a Coma-like cluster of mass $M_{\text{vir}} = (1.2 \pm 0.3) \times 10^{15}~h^{-1}~{\rm M}_{\odot}$, where $h = \frac{\rm H_{0}}{100 \rm km s^{-1} Mpc^{-1}}$. \\
    \item [CC2.2:] This protocluster was spectroscopically confirmed by \citet{Darvish_2020}, following up a large relative overdensity at this location \citep{Darvish_2017}. \citet{Darvish_2020} estimates a virial mass of $M_{\rm vir} = (1-2) \times 10^{14} \rm M_{\odot}$ for this structure at its observed redshift, $z \approx 2.2$.
    In the constrained simulations, a cluster is found at this location 42 out of 50 times, with an associated mass of $M_{\text{vir}} = (4.2 \pm 1.9) \times 10^{14}~h^{-1}~{\rm M}_{\odot}$ by $z = 0$. \\
    \item [Hyperion (1-7):] The Hyperion protoclusters were individually found by several studies \citep{Lee_2016, Diener_2015, Chiang_2015, Casey_2015, Wang_2016} before a connection between them was made by \citet{Cucciati_2018}, which found the system had an estimated total mass of $4.8 \times 10^{15} M_{\odot}$ over a volume of $\sim$ 60 x 60 x 150 cMpc$^3$ at $z \sim 2.45$. It was originally hypothesized that this collection of seven density peaks will evolve into a super-cluster by $z = 0$, with the various peaks virializing by redshift $z \sim 0.8 - 1.6$ \citep{Cucciati_2018}.  However, the constrained simulations show that by $z = 0$, four virialized clusters emerge and form a filamentary group of clusters with a total mass of $M_{\text{vir}} = (2.5 \pm 0.5) \times 10^{15}~h^{-1}~{\rm M}_{\odot}$ spanning $(65 \pm 10)~h^{-1}$~Mpc. This projected structure is expected to be similar in spatial extent and mass to the Coma/A1367 filament in the local Universe \citep{1984A&A...138...85F}.\\
    \item[COSTCO Protoclusters:] The COnstrained Simulations of The COsmos field (COSTCO) protoclusters are a set of protoclusters found purely through the constrained simulations suite presented in \citet{Ata_2022}. While they do not have strong overdensities throughout $2 < z < 2.52$, they are extended structures that collapse into Virgo-like clusters ($\sim 10^{14.5}~{\rm M}_{\odot}$) by $z = 0$.  
    COSTCO J100026.4+020940 has previously been identified as an overdensity \citep{Lee_2016}.  Recently, \citet{Dong_2023} noted that the large scale gas associated with this protocluster has been heated far higher than expected.
    COSTCO J095945.1+020528 is found to collapse into a cluster only 27 out of 50 times, though in 40 of those simulations it still results in a substantial overdensity at $z=0$.  COSTCO J095945.1+020528 is just south of Hyperion and might become a substructure of it. Tidal disruptions by Hyperion may be the reason why this does not collapse into an independent virialized structure in all cases \citep{Ata_2022}.
\end{description}

We identify candidate cluster members by selecting all galaxies within a projected radius $dR$ and a photometric redshift range $dz$.  For sufficiently large $dR$, these volumes for neighbouring clusters will partially overlap.  Since the $dz$ must be large enough to accommodate the significant photometric redshift uncertainties, the volume will be much larger than the physical volume occupied by the cluster, and will include many non-cluster members.  These must be corrected statistically, which requires an accurate volume calculation. This is done using a Monte Carlo approach. We take a `box' of Cartesian space surrounding the clusters, and uniformly populate it with $10^7$ points. We first remove all the points outside the UltraVISTA rectangle \citep{McCracken_2012}, or in masked regions. We then take the fraction of points inside protocluster cylinders and multiply it by the volume of the box to measure the volume of the protoclusters. For example, for our fiducial cluster volume (see Table \ref{tab:selection_quantities}) which has properties $dR = 1$ Mpc and $dz = 0.2$, there are $\sim$ 7750 points inside the protocluster volume. This gives us a volume of $\sim$ 192,000 $\pm$ 2,200 Mpc$^3$ assuming a Poisson counting error.
This precision of $\sim 1$ \% is sufficient that it does not dominate our error budget.  

\section{Stellar mass functions} \label{Analysis}

\subsection{Methodology}\label{sec:methods}

To determine the observed number densities, we bin our data by mass and weight each bin by dividing the count by bin size and volume corresponding to the region in question. We take the uncertainty of this to be simply the square root of each count for the respective bins divided by the associated volume. 

To fit the unbinned data above the stellar mass limit, we closely follow the Parametric Maximum-Likelihood method \citep{1979ApJ...232..352S}. We will fit our data with a double \citep{1976ApJ...203..297S} function, as defined in \citet{2008MNRAS.388..945B}, 
in terms of $\mathcal{M}=\log_{10}(M_{\ast}/{\rm M}_{\odot})$:

\begin{multline} \label{eq:DS}
    \phi(\mathcal{M}) = \text{ln}(10) \cdot \text{exp}(-10^{(\mathcal{M} - \mathcal{M}^{*})}) 
    \cdot 10^{(\mathcal{M} - \mathcal{M}^{*})} \\
    \cdot \left[\phi^{*}_{1}\cdot10^{(\mathcal{M} - \mathcal{M}^{*})\alpha_1} + \phi^{*}_{2}\cdot10^{(\mathcal{M} - \mathcal{M}^{*})\alpha_2}\right]
\end{multline}
where $\phi(\mathcal{M})$ is the number of galaxies per Mpc$^{3}$ per dex and $\mathcal{M^{*}} = \text{log}(M^{\ast}/{\rm M}_{\odot})$ is the characteristic mass.  The parameters $\alpha_1$ and $\alpha_2$ are the high- and low-mass slopes, respectively, with corresponding normalizations $\phi_1^{*}$ and $\phi_2^{*}$. This is effectively adding together two Schechter functions with the same $\mathcal{M}^{*}$.
We then assign a probability to each galaxy, as first presented in \cite{1986AJ.....91..697O} and \cite{1986ApJ...308...10M}:

\begin{equation} \label{eq:prob}
    p_i \equiv p(\mathcal{M}_i) = \frac{\phi(\mathcal{M}_i)}{\int^{\infty}_{\mathcal{M}_{\rm lim}} \phi(\mathcal{M}) d\mathcal{M}}.
\end{equation}
The likelihood of any given model is defined as the sum of the logarithms of the individual probabilities for each galaxy considered. 
We determine the parameters $\mathcal{M}^*$,  $\alpha_1$, $\alpha_2$ and the ratio $\phi^{*}_{2}/\phi^{*}_{1}$ via an MCMC chain. The overall normalization $\phi^{*}_{1}$ is set by forcing the integral of the function above the mass limit to equal the number density of galaxies in the sample. 
We set our uniform priors to be $\alpha_1$ $\subseteq$ [-3, 1.5], $\alpha_2 \subseteq$ [-3, -1], $\mathcal{M}^* \subseteq$ [9.5, 12] and $\phi^{*}_{2}/\phi^{*}_{1}$ $\subseteq$ [0, 0.5]. 
While these priors for $\alpha_1$ and $\mathcal{M}^{\ast}$ are broad and uninformed, the choice of $\alpha_2$ and $\phi^{*}_{2}/\phi^{*}_{1}$ are specifically motivated to ensure the second component corresponds to any low mass upturn, rather than other possible deviations from a single Schechter function at high mass.

To measure the protocluster SMFs, we measure the SMF in a volume centered on the protoclusters (see Section~\ref{sec:contrast}). However, as described in Section~\ref{sec:contrast}, this region is heavily contaminated with field galaxies. To accommodate this, we adjust Equation~\ref{eq:DS}:

\begin{equation} \label{eq:model}
    \phi(\mathcal{M}) = \phi_{f}(\mathcal{M}) + \phi_{c}(\mathcal{M}),
\end{equation}
where the $f$ and $c$ subscripts are for the field and cluster contributions to the protocluster volume respectively.  Both $\phi_{f}(\mathcal{M})$ and $\phi_{c}(\mathcal{M})$ are double Schechter functions as in Equation~\ref{eq:DS}. We measure $\phi_{f}$ for the full field sample (see next section). We then can measure $\phi_{c}$ by fitting for Equation~\ref{eq:model}, determining the parameters $\mathcal{M}^*_{c}$,  $\alpha_{1,c}$, $\alpha_{2,c}$ and the ratios $\phi^{*}_{2, c}/\phi^{*}_{1, c}$ and $\phi^{*}_{1, f}/\phi^{*}_{1, c}$. We then determine $\phi^{*}_{1, c}$ in the same way we set $\phi^{*}_{1}$.

This allows us to determine intrinsic protocluster SMF to each (unbinned) population. We also consider the binned data for each population, measured by subtracting the field component in each bin. This is described further in Section~\ref{sec:PSMFs}.

\subsection{Field Stellar Mass Function and Comparison to Literature}

\begin{figure}
\centering
\includegraphics[width=1\linewidth]{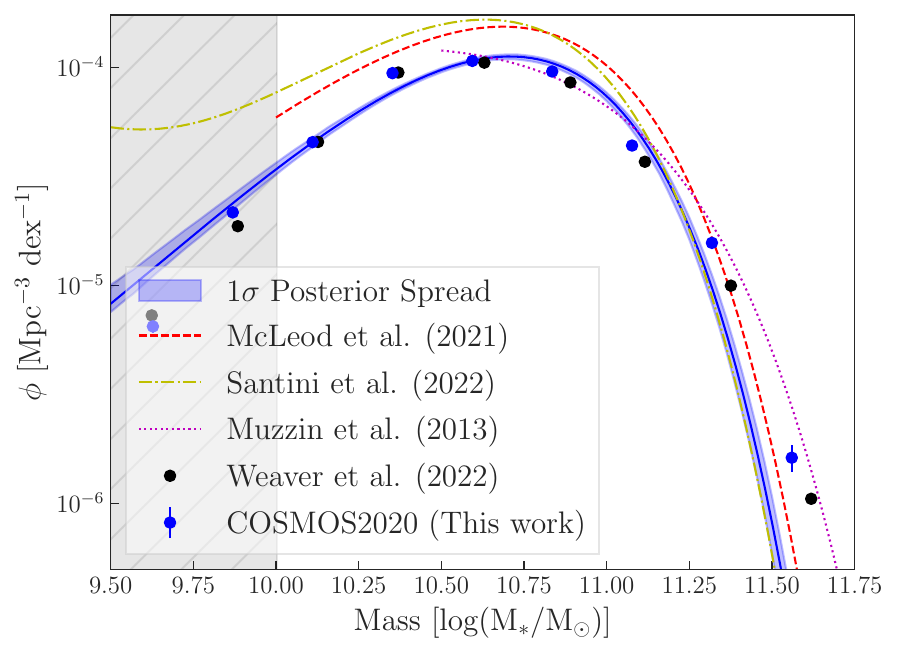}
\caption{We compare the total field SMF at $2<z<2.5$ (solid line) with measurements from  
\citet{McLeod_2021}, \citet{Santini_2022}, \citet{Muzzin_2013} and \citetalias{Weaver_SMF}.  The results are generally consistent with one another, though there is some variation at the high-mass end.
Note there is a small difference in redshift ranges considered, as   \citet{McLeod_2021} and \citet{Santini_2022} are presented for data between $2.25 < z < 2.75$, while \citet{Muzzin_2013}, \citetalias{Weaver_SMF} and this work are between $2 < z < 2.5$. See Appendix~\ref{app-field} for a presentation using different redshift bins.  The hatched region represents the mass range between our conservative and optimistic mass completeness limits, as discussed in Section \ref{mass completeness}.}
\label{fig:rel_lit_comp}
\end{figure}

In Fig. \ref{fig:rel_lit_comp}, we show the derived field SMFs observed in our subsample of the COSMOS2020 survey at $2<z<2.5$. This definition of the `field' is simply everything in our sample, which includes both low- and high-density regions. We recover closely the result presented in \citetalias{Weaver_2022}, as expected since we are using the same catalogue.  This also agrees reasonably well with the total SMFs presented in \cite{Muzzin_2013}, \cite{McLeod_2021} and \cite{Santini_2022}  at a similar redshift.  Our results show some sensitivity to the redshift range of the field sample, which has been chosen to correspond well to the redshift distribution of our protocluster sample, as described in Appendix~\ref{Field Choice}.

\begin{figure}
\centering
\includegraphics[width=1\linewidth]{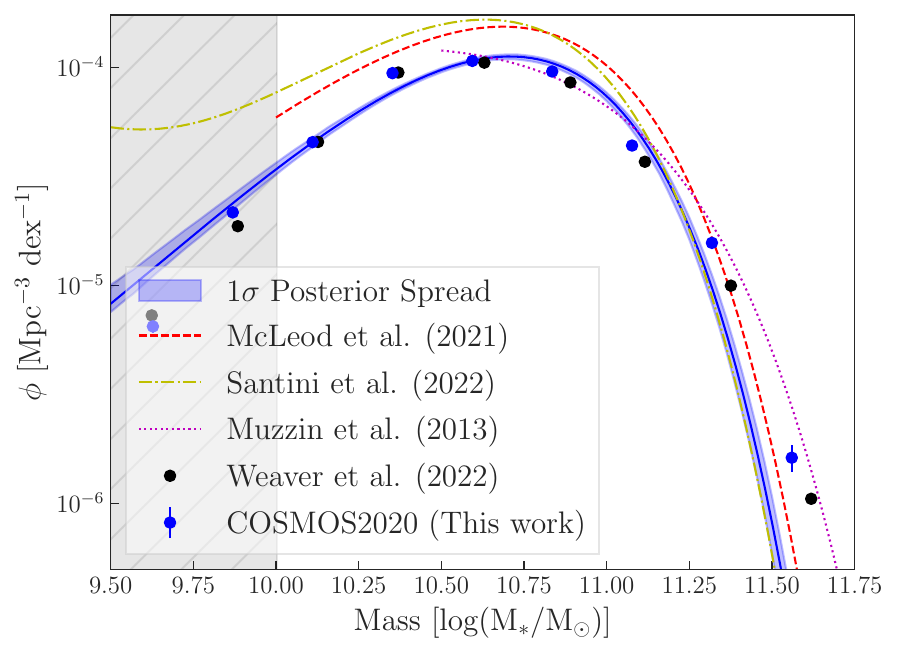}
\caption{We compare the quiescent field galaxy SMF at $2<z<2.5$ (solid line) for our selected sample with those from \citetalias{Weaver_SMF}, \citet{McLeod_2021}, \citet{Santini_2022} and \citet{Muzzin_2013}. Note that \citet{Muzzin_2013} and \citet{Santini_2022} use $UVJ$ colours to define quiescent galaxies, while the others (including our work) use ${\rm NUV}rJ$.  Also, the \citet{McLeod_2021} and \citet{Santini_2022} results are for a different redshift range of $2.25<z<2.75$.  See Appendix~\ref{app-field} for a presentation using different redshift bins and colour selections.  The hatched region represents the mass range between our conservative and optimistic mass completeness limits, as discussed in Section \ref{mass completeness}.}
\label{fig:rel_lit_comp2}
\end{figure}

We make the same comparison for the quiescent population in Fig. \ref{fig:rel_lit_comp2}. While our results are in good agreement with \cite{Muzzin_2013} and \citetalias{Weaver_SMF}, both \cite{McLeod_2021} and \cite{Santini_2022}  find significantly larger numbers of quiescent galaxies, especially at low stellar masses. Although the redshift ranges do not match exactly, we show in Appendix~\ref{app-field}, Figure~\ref{fig:lit_comp2}, that this does not account for the difference.  Differences in the definition of quiescent galaxies are also unlikely to be the explanation.  While \citet{Santini_2022} and \citet{McLeod_2021} use $UVJ$ colours, rather than the ${\rm NUV}rJ$ selection that we adopt, we show in Figures \ref{fig:lit_comp2} \& \ref{fig:UVJ_lit_comp} that this choice does not make a significant difference to the quiescent SMF that we derive \citep[see also][]{gould2023cosmos2020}.  Furthermore, \citet{Muzzin_2013} also use a $UVJ$ definition, and their result is similar to ours.  Cosmic variance is estimated to account for an uncertainty of only $\sim 20\%$ for the \cite{Santini_2022} and \cite{McLeod_2021} samples, which is small relative to the $\gtrsim 70$ per cent difference between SMFs. We have checked, using three independent samples of $\sim 1000$ arcmin$^2$ \citep[corresponding to the survey area in ][]{Santini_2022} within our sample of COSMOS data, that the cosmic variance for the quiescent population is not significantly different than that of the total over most of the mass range.
It is possible that the difference lies in the stellar mass or redshift estimates, though all studies use similar methods (e.g. parametric star formation histories, dust law, etc.). 
Thus, the larger population of quiescent galaxies in \citet{McLeod_2021} and \citet{Santini_2022} remains unexplained.  If it is due to an incompleteness in COSMOS2020, we would expect that to affect our target protocluster volumes (which are only modestly overdense, see Section~\ref{sec:contrast}) similarly to the general field, in which case any impact on our conclusions based on the comparison of these samples will be small.

\section{Results} \label{Results}

\subsection{Protocluster contrast}\label{sec:contrast}
\begin{figure}
\centering
\includegraphics[width=1\linewidth]{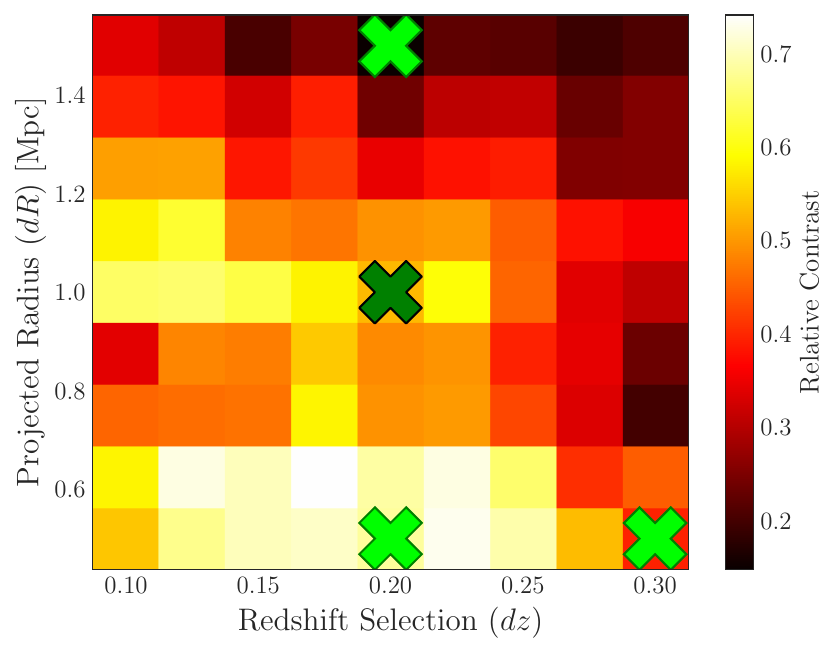}
\caption{The relative contrast is shown as a function of cluster radius $dR$ and redshift selection $dz$, with our selections (see Table \ref{tab:selection_quantities}) indicated by the crosses. Our fiducial sample at $dR=1$ Mpc and $dz=0.2$ has a contrast of $\sim 50$ per cent, and is indicated by the darker cross.  A higher contrast is found at smaller $dR$ and $dz$, at the expense of a less complete sample. Note that $dR$ is in physical units, not comoving.}
\label{fig:overdense_colourmesh}
\end{figure}

Due to the large selection volume necessitated by the photometric redshift uncertainties (see Section~\ref{membership}), significant field contamination is expected. To quantify this, we measure the contrast of our protocluster sample relative to the field. We calculate the contrast by finding the total number density of galaxies with log$_{10}(M_{\ast}/M_{\odot}) > 10.5$ in the protocluster selection volume, and subtract the field contribution within that volume from the global SMF. We then divide this quantity by the overall field density to get the relative contrast. In Fig. \ref{fig:overdense_colourmesh} we show how this contrast depends on the choice of $dR$ and $dz$.
The contrast of our protocluster sample relative to a random field is low, $\lesssim 80$ per cent.  The physical overdensities are likely much higher.  For example, if we assume that the protocluster galaxies are contained in a sphere of radius $dR=1$Mpc, this contrast corresponds to a  physical overdensity of $\sim$1600.

Our fiducial protocluster selection of $dR=1$ Mpc and $dz=0.2$ is physically motivated.  The radial extent is chosen to correspond approximately to the virial radius of descendent clusters at $z\sim 1.3$, as discussed further in Section~\ref{sec:evolution}.  The $dz=0.2$ selection is chosen to correspond to the 95th percentile of photometric redshift error for star-forming galaxies, and is still close to the peak contrast shown in Figure~\ref{fig:overdense_colourmesh}. 
In addition to the fiducial sample, we also consider a "Core" sample restricted to $dR=0.5$ Mpc.  The contrast of this sample is higher, at the cost of a greatly reduced sample size (see Table \ref{tab:selection_quantities}).  At the other extreme, we consider a "Wide" sample with $dR=1.5$ Mpc.   While it is known that protocluster structures can extend to even larger distances \citep{Muldrew_2015, Chiang_2013, Contini+16}, field contamination dominates in such a volume, making a comparable analysis impractical. 
Finally we consider a "Core-complete" sample with $dR=0.5$ Mpc and $dz=0.3$\footnote{To create this sample, we apply the same cuts described in section~\ref{Sample Selection}, but draw from $1.7<z<2.8$ instead. We still restrict our analysis to $M_{\ast} \geq 10^{9.5} M_{\odot}$.}. The larger redshift selection improves the completeness of the sample for quiescent galaxies.   The different samples are listed in Table~\ref{tab:selection_quantities}, together with
the number of quiescent and star-forming galaxies, as defined in Section~\ref{NUVrJ}. 
The NUVrJ colour distributions of each sample are shown in Appendix~\ref{Backgroun Subs}, Figure~\ref{fig:mult_NUVrJ}.

\begin{table}
\begin{tabular}{llllll}
\hline
Selection & Alias       &  $dR$, $dz$       & N$_{\rm q}$ & N$_{\rm sf}$ & N$_{\rm tot}$ \\ \hline
A & Fiducial &      1.0, 0.2 & 32 & 431 & 463  \\
B & Core     &      0.5, 0.2 & 10 & 118 & 128  \\ 
C & Wide     &      1.5, 0.2 & 51 & 833 & 884  \\ 
D & Core-Complete & 0.5, 0.3 & 13 & 162 & 175 \\ \hline
\end{tabular}
\caption{The number of quiescent, star-forming and total galaxies in each selection of $dR$ and $dz$. 
Most of the analysis in this paper is based on the fiducial sample A, as a good balance between completeness and purity.  Selected results for the other samples are provided in Appendix~\ref{Backgroun Subs} and Table\ref{tab:vals}. }
\label{tab:selection_quantities}
\end{table}

\subsection{Protocluster Stellar Mass Functions}\label{sec:PSMFs}

\begin{figure*}
\centering
\includegraphics[width=\linewidth, trim = 75 0 75 0,clip]{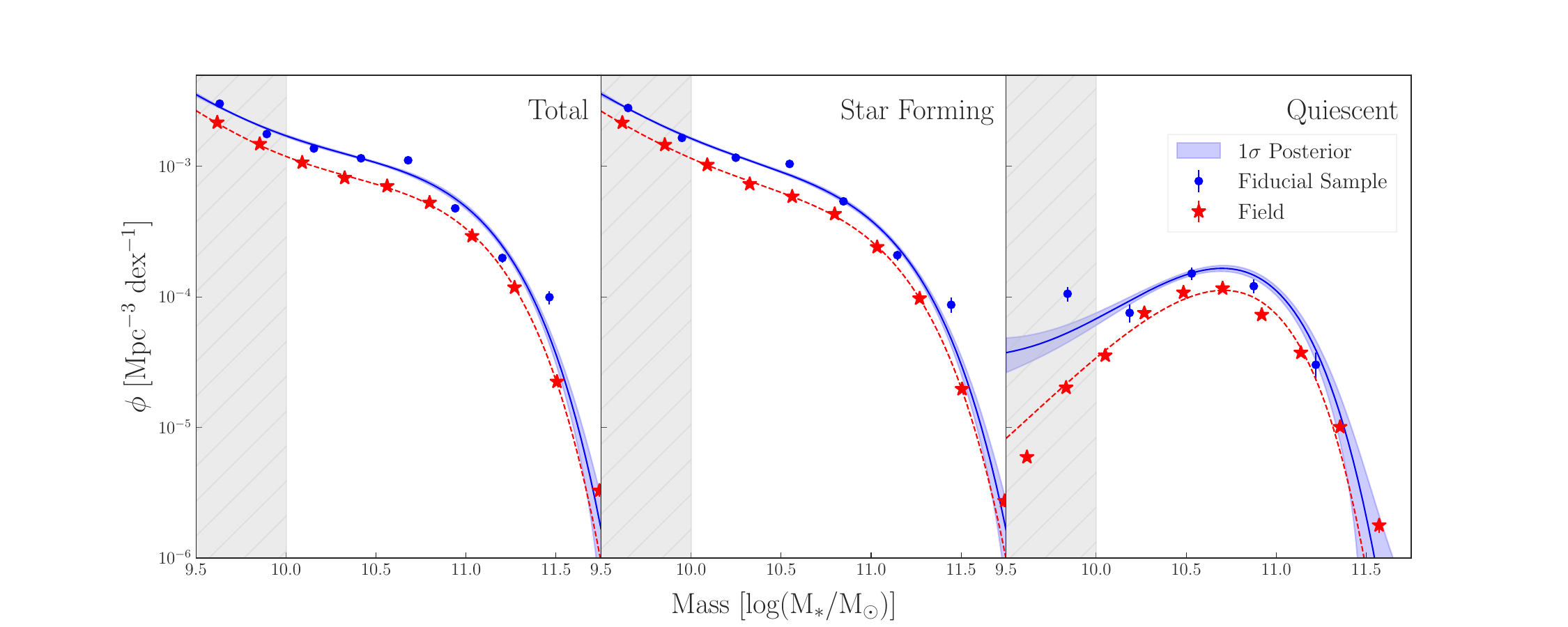}
\caption{The SMFs within the fiducial volume selection A containing the protoclusters,  $dR = 1$ Mpc and $dz = 0.2$.  The shaded region shows the $1\sigma$ uncertainty about the best-fit double Schechter function, which is fit to the unbinned data. The contrast relative to the general field is significant, for the total and star-forming population, at $M_{\ast}<10^{11}M_{\odot}$. There is evidence for an excess of quiescent galaxies at low stellar masses in the protocluster volumes, relative to the field. The hatched region represents the mass range between our conservative and optimistic mass completeness limits, as discussed in Section \ref{mass completeness}.}
\label{fig:SMF_1_02_all}
\end{figure*}

In Fig. \ref{fig:SMF_1_02_all}, we show the SMFs for our fiducial selection. Both the total and star-forming and quiescent populations show a significant overdensity relative to the average field, for log$_{10}(M_{\ast}/M_{\odot})\lesssim 11$. However, while the shape of the star-forming galaxy SMF is similar to that of the field, the  SMF for quiescent galaxies in this volume has an overall flatter shape, indicating relatively more low-mass galaxies than observed in the field.  Note that here we are showing show the fit for $\phi(\mathcal{M}) = \phi_{f}(\mathcal{M}) + \phi_{c}(\mathcal{M})$, as shown in Equation~\ref{eq:model} and discussed in Section~\ref{sec:methods}.

\begin{figure*}
\centering
\includegraphics[width=\linewidth, trim= 0 0 0 0, clip]{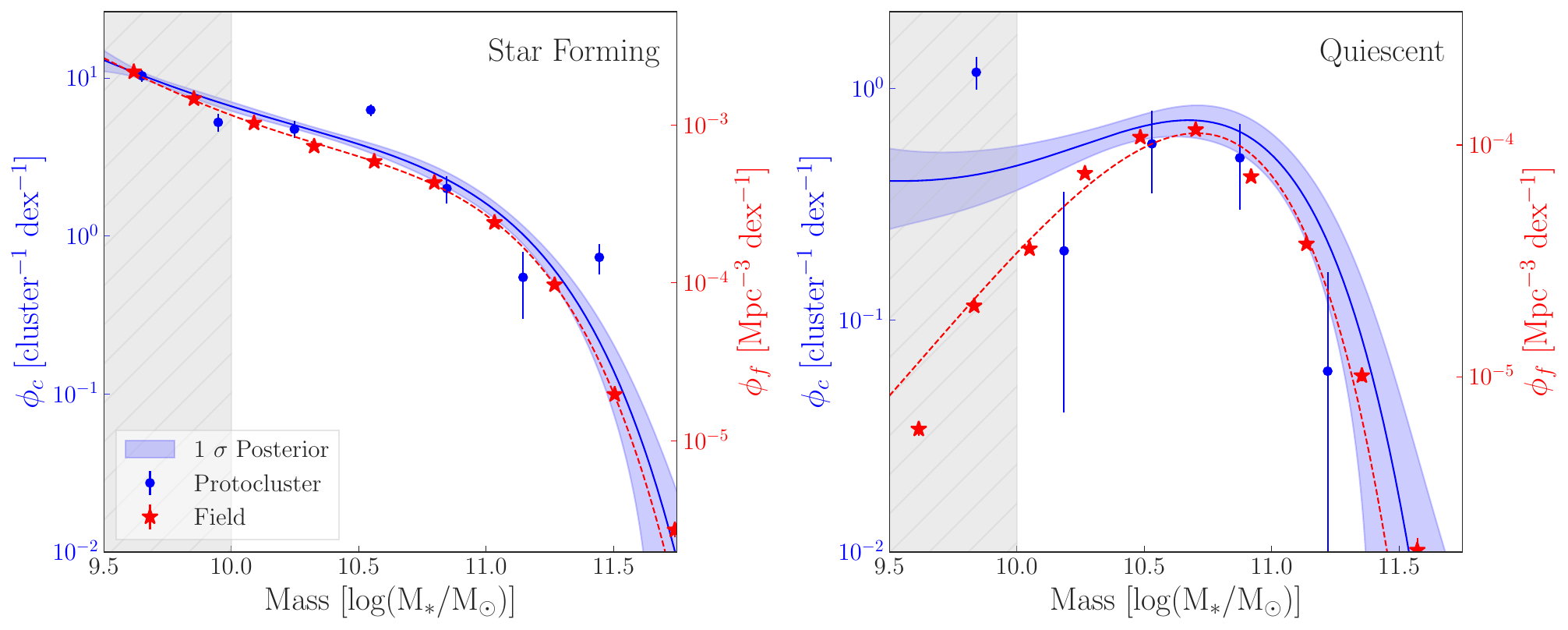}
\caption{The intrinsic SMFs for protocluster galaxies in the fiducial sample (A, Table \ref{tab:selection_quantities}) are shown for star-forming (left) and quiescent (right) galaxies.  The binned measurements are shown as the blue points, with the blue shaded region representing the difference in the fits to the unbinned data. The SMF of the protocluster is presented in units of $\text{dex}^{-1} \text{cluster}^{-1}$ (blue, left axis) since the physical volume occupied by the protocluster is unknown. For comparison we show the field SMF in red, and the associated y-axis range (red, on the right side of the panel, in units of $\text{dex}^{-1} \text{Mpc}^{-3}$) has been chosen to facilitate comparison of the shapes of the two SMFs such that the field and cluster align near $M^\ast$. 
The shape of the SMFs of the star-forming population matches the shape of the field SMF well, given the uncertainties. However, the quiescent galaxy SMF has a qualitatively different shape from the field.  While the number of quiescent galaxies in the field decreases monotonically towards lower masses, in the protocluster an upturn is seen, leading to a relative excess of quiescent galaxies at low mass.
Similar plots for the other volume selections of Table~\ref{tab:selection_quantities} are shown in Appendix~\ref{Backgroun Subs}. The grey hatched region represents the mass range between our conservative and optimistic mass completeness limits, as discussed in Section \ref{mass completeness}.
}
\label{fig:FSub_SMF_1_02_all}
\end{figure*}

As discussed in Section~\ref{sec:methods}, we also measure the intrinsic protocluster SMF, $\phi_c$. The results for our fiducial sample are shown in Fig. \ref{fig:FSub_SMF_1_02_all}, in units of dex$^{-1}$ cluster$^{-1}$ (left, blue axis). Note this normalization is per cluster rather than per unit volume, since the physical volume occupied by the overdensity within our large cylinders is unknown. This is done by multiplying the SMF by the protocluster selection volume and then dividing by the number of clusters. For comparison we plot the corresponding field SMFs, here in units of dex$^{-1}$ Mpc$^{-3}$ (right, red axis).   In this representation, the relative normalization of the cluster and field curves has no meaning since the units are not the same.  To facilitate comparison of the shapes, we arbitrarily adjust the axis limits. This illustrates how the shape of the dominant star-forming population in the protocluster sample is indistinguishable from that of the field, with a monotonically increasing number of galaxies toward lower stellar mass.  However, the shape of the quiescent SMF in the protocluster is qualitatively different from that in the field.  While both the field and protocluster SMF peak at log$_{10}(M_{\ast}/M_\odot)\approx 10.75$, the protocluster SMF does not drop off, instead showing signs of an upturn, leading to a relative excess at log$_{10}(M_{\ast}/M_{\odot}) < 10.5$.

\begin{figure}
    \centering
    \includegraphics[width=\linewidth, trim = 0 0 0 0, clip]{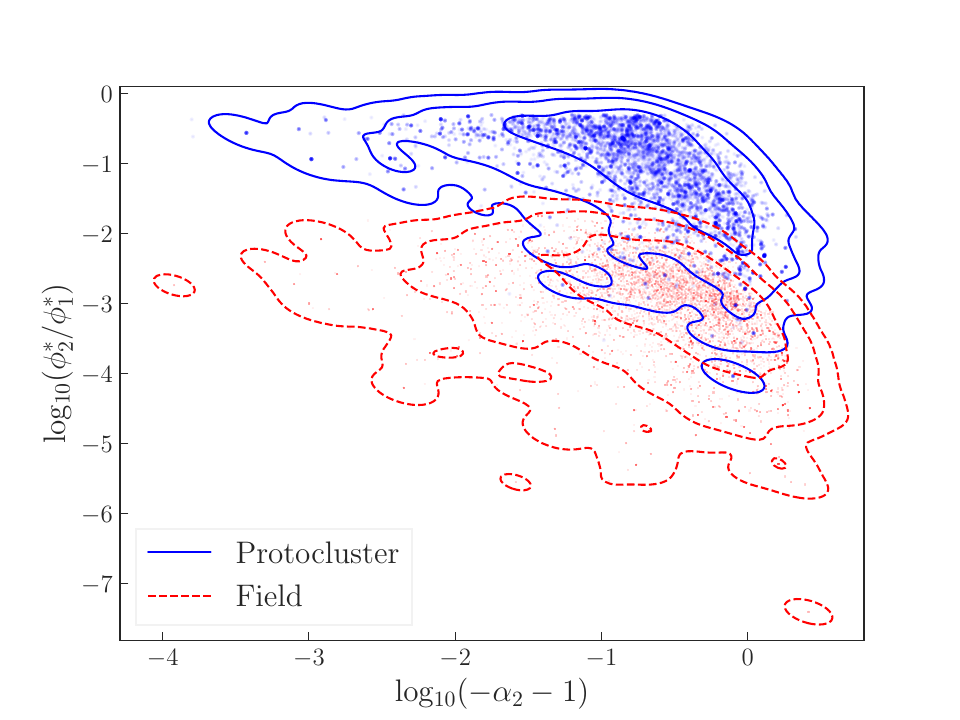}
    \caption{The distribution of the log of the ratio $\phi^{\ast}_{2}/\phi^{\ast}_{1}$ and parameter log($-\alpha_{2} - 1$) for both the protocluster (blue) and the field (red, dashed) quiescent populations. The curves show the 1, 2 and 3$\sigma$ contour levels.
    This parameter combination in the protocluster differs from that of the field at the $\sim 2\sigma$ level.
    }
    \label{fig:phi vs alpha}
\end{figure}

To quantify the significance of the difference in the quiescent galaxy SMFs between the protocluster and the field, we show the confidence intervals of the ratio $\phi^{\ast}_{2}/\phi^{\ast}_{1}$ and parameter $\alpha_{2}$ for both populations in Fig~\ref{fig:phi vs alpha}. These parameters characterize the low mass upturn in the SMF, where we observe qualitatively different SMFs in the protocluster and field samples.  Each distribution is generated from the MCMC chain, measured as described in section~\ref{sec:methods}. For display clarity we plot log($\phi^{\ast}_{2}/\phi^{\ast}_{1}$) and log($-\alpha_{2} - 1$).
This shows a $\sim 2\sigma$ difference in this parameter combination between the protoclusters and the field.  We conclude, therefore, that the difference in shapes at the low mass end is intriguing but not statistically significant.

\begin{figure*}
\centering
\includegraphics[width=\linewidth]{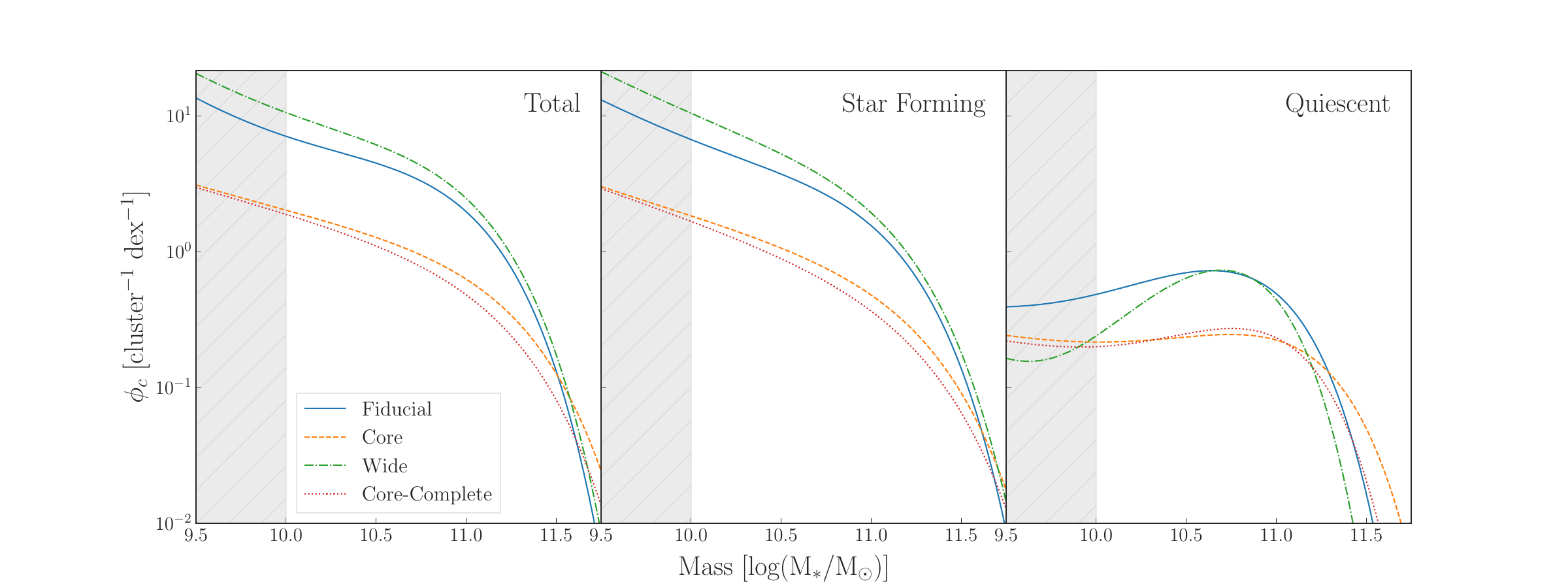} 
\caption{The intrinsic SMFs for each selection volume (Table~\ref{tab:selection_quantities}). To clearly show the qualitative differences between volume selections, we omit the uncertainty ranges on these figures.  
The hatched region represents the mass range between our conservative and optimistic mass completeness limits, as discussed in Section \ref{mass completeness}. Individual results for each selection are shown in Appendix \ref{Backgroun Subs}.
}
\label{fig:All_intrinsic_selections}
\end{figure*}

We show how the intrinsic protocluster SMFs depend on our different selections in Figure~\ref{fig:All_intrinsic_selections}. To allow a clear comparison of the relative shapes on a single plot, we do not show the uncertainty ranges, which are especially large for the two Core samples.
The best-fit parameters for all double Schechter function fits are provided in Table \ref{tab:vals}. See Appendix~\ref{Backgroun Subs} for more details and uncertainty ranges for each sample.

\section{Discussion}\label{Discussion}

\subsection{Excess of Low-Mass Quiescent Galaxies} \label{low mass excess}
\begin{figure}
\centering
\includegraphics[width=1\linewidth]{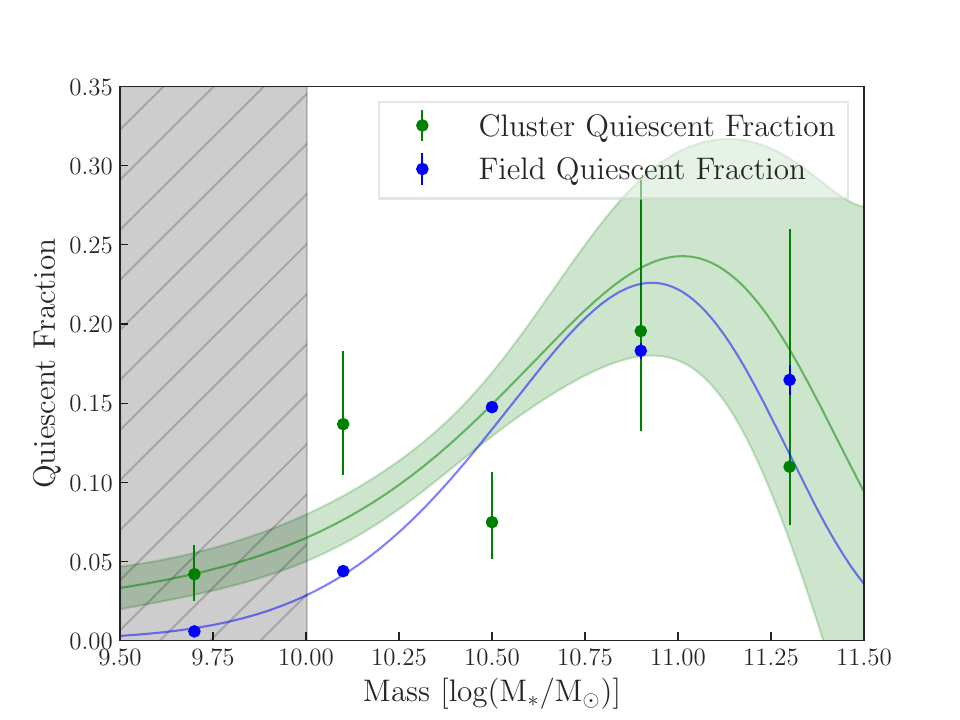} \label{fig:q frac}
\caption{The quiescent fraction for galaxies within our fiducial volume selection (A). The field and cluster are statistically indistinguishable for log$_{10}(M_{\ast}/M_{\odot}) > 10.5$. At lower masses, the cluster sample shows a small excess of quiescent galaxies, though the statistical significance is low.  Error bars on the binned data represent the 1$\sigma$ binomial confidence interval.  The shaded region is derived from the double-Schechter function fits to the unbinned protocluster and field data shown in Figure~\ref{fig:SMF_1_02_all}.}
\label{fig:Fsub_QF}
\end{figure}

Figures ~\ref{fig:SMF_1_02_all} and ~\ref{fig:FSub_SMF_1_02_all} show a moderately significant excess of low-mass quiescent galaxies within the protocluster regions.  
To explore this further, in Figure~\ref{fig:Fsub_QF} we show the quiescent fraction for our fiducial protocluster sample ($dR$, $dz$ = 1 Mpc, 0.2), after field subtraction.  The fraction is generally quite low, with a 1$\sigma$ upper limit of $\approx 0.15$ for stellar masses with log$_{10}(M_{\ast}/M_{\odot}) < 10.75$.
For most of the stellar mass range, and certainly for log$_{10}(M_{\ast}/M_{\odot}) > 10.25$, the field and protocluster population have quiescent fractions that are fully consistent with one another, within the substantial 1$\sigma$ uncertainties. At lower masses we find evidence for a small excess in quiescent fraction, though the statistical significance is not high enough to make strong claims, and larger samples will be required to confirm this. Almost all of these low-mass quiescent galaxies are from the ZFOURGE/ZFIRE protocluster (see Figure~\ref{fig:fsub_ZFORGE}),
which is the most massive protocluster in our sample.  We note also that the apparent excess is in the mass regime where the sample may suffer some incompleteness, though we would expect this incompleteness to affect the protocluster and field samples similarly.

\subsection{Comparison with other protocluster literature}
As we discuss further in the following subsection, the lack of a dominant quiescent population in these protoclusters is surprising.  This result also contrasts with some recent claims for quiescent populations in protoclusters at a similar redshift.  In an analysis of the cluster QO-1000 at $z=2.77$, \citet{Ito_2023} found 14 quiescent galaxies above mass $\log_{10}{(M_{\ast}/{\rm M}_{\odot}}$) $> 10.5$, with a number density excess of 4.2$\sigma$ and a quiescent fraction of $0.34 \pm 0.11$.  Similarly, \citet{Ando_2020} analyze 75 protocluster "cores" at $1.5<z<3$ in the COSMOS field, using pairs of massive galaxies as tracers.  They find a quiescent fraction of $0.17\pm 0.04$, three times larger than the field, for log$(M_{\ast}/{\rm M}_{\odot}) > 9$.  We note, though, that this sample is dominated by systems at $z<2$. At somewhat higher redshift $z\approx 3.4$, \citet{McConachie_2022} discovered a couple of protoclusters, one of which has a very high quiescent fraction of $0.73^{+0.27}_{-0.17}$ among the most massive galaxies log$(M_{\ast}/{\rm M}_{\odot}) > 11.3$.

To compare with these and other studies, we consider the Quenched Fraction Excess \citep[QFE,][]{vandenBosh2008satQuenching,Wetzel_2012, Bahe+17, van_der_Burg_2020}. This quantity shows the fraction of galaxies that are quenched in the high-density protocluster region, but would expected to be star-forming in the field. This is given by:

\begin{equation}\label{eqn:QFE}
    {\rm QFE} = \frac{f_{\rm q, clus} - f_{\rm q, field}}{1 - f_{\rm q, field}},
\end{equation}
where $f_{\rm q, clus}$ and $f_{\rm q, field}$ are the cluster and field quenched fractions, respectively.  Since $f_{\rm q, field}$ is quite small at this redshift, $\lesssim 0.2$, in practice this is not very different from $f_{\rm q, clus}$.

\begin{figure*}
\centering
\includegraphics[width=1\linewidth]{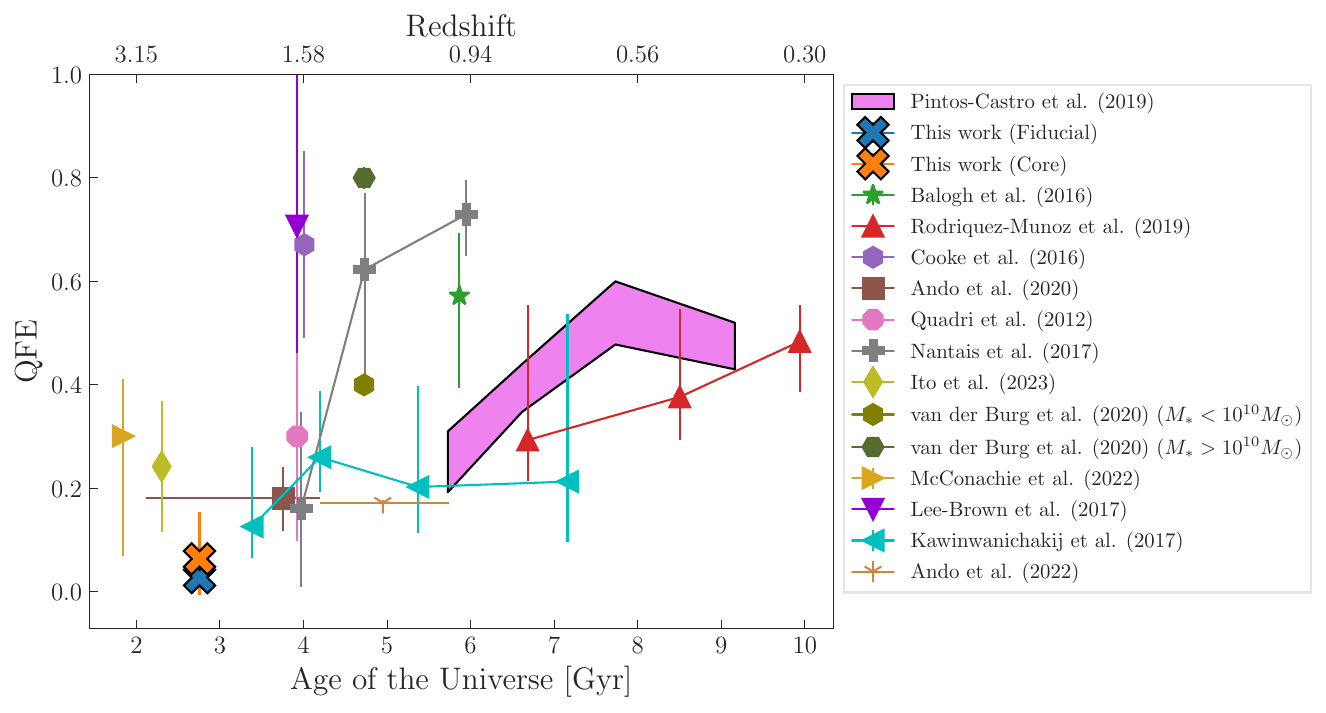}
\cprotect\caption{The QFE (Equation~\ref{eqn:QFE}) for clusters and protoclusters in our work and the literature as a function of redshift. The mass and cluster-centric radius limits for each work are summarized in Table \ref{tab:QFE mass lims}.  Most are representative of the population with $M_{\ast}>10^{10} M_{\odot}$, with major exceptions noted in the legend. 
The blue and orange crosses represent our measurements, for the Fiducial and Core samples, respectively, for $\log_{10}{M_{\ast}/\mbox{M}_\odot}>9.5$.
}
\label{fig:QFE lit}
\end{figure*}

We calculate the QFE for our sample and compare it to that of other works in the literature in Fig. \ref{fig:QFE lit}. We do this for both our fiducial (blue cross) and core (orange cross) samples, considering all galaxies $M_{\ast} > 10^{9.5} M_{\odot}$. We measure a QFE of 0.03$^{+0.04}_{-0.03}$ in the Fiducial sample, and 0.06$^{+0.09}_{-0.07}$ for the Core. While low, this is within $\lesssim 2\sigma$ of other high redshift studies such as \citet{McConachie_2022}, \citet{Ito_2023} and \citet{Ando_2020}.

We noted previously that most of the low-mass quiescent galaxies in our sample come from the most massive system, ZFOURGE/ZFIRE.  Considering only this protocluster, 
using our fiducial parameters $dR, dz = 1, 0.2$, 
we measure a QFE of 0.13$^{+0.12}_{-0.11}$ over the mass range $9.5 \leq \text{log}_{10}(M_{\ast}/M_{\odot} \leq 11.5)$. Over the same mass range, we measure a QFE of 0.00$^{+0.04}_{-0.03}$ for the rest of our protoclusters. These two measurements are consistent within 1$\sigma$, but hint at a halo-mass dependence that will require larger samples to confirm. 


Fig. \ref{fig:QFE lit} does not present a very clear trend, and the interpretation is complicated by the dependence of QFE on stellar mass, local density or clustercentric radius, and possibly halo mass.  There is some indication that a modest quiescent excess is already present in some protoclusters at $2<z<3$, and that there is significant evolution toward higher QFE in cluster cores already by $z=1.6$, only $\sim 2$ Gyr later.  We explore this further in the following section.


\begin{table*} 
\begin{tabular}{@{}lll@{}}
\toprule
\textbf{Work}               & \textbf{Mass Limit {[}log$_{10}(\frac{M_{\ast}}{M_{\odot}})${]}} & \textbf{Clustercentric Radius}\\ \midrule
\cite{Ando_2020}            & 9.5                                      & 0.5 Mpc                        \\ 
\cite{Ando_2022}            & 10                                       & 0.5 Mpc                         \\
\cite{Balogh_2016}          & 10.5                                     & 1 Mpc                            \\
\cite{Cooke_2016}           & 10.7                                     & $\sim 1$ Mpc                      \\
\cite{Ito_2023}             & 10.3                                     & 1 Mpc                              \\ 
\cite{K+17}                 & 10.2                                     &                                     \\
\cite{LB+17}                & 10.2                                     & 0.6 Mpc                              \\ 
\cite{McConachie_2022}      & 10.5                                     & 2.3 Mpc                               \\ 
\cite{Nantais_2016}         & 10.3                                     & 1 Mpc                                  \\
\cite{PC+19}                & 10.2                                     & 0.4 R/R$_{200}$                         \\
\cite{Quadri_2011}          & 10                                       & 0.4 Mpc                                  \\ 
\cite{Rodr_guez_Mu_oz_2019} & 10                                       & 0.2 Mpc                                   \\
\cite{van_der_Burg_2020}    & 9.5                                      & 1 Mpc                                      \\ \bottomrule
\end{tabular}
\cprotect\caption{The mass limits and clustercentric radius for each work shown in Fig. \ref{fig:QFE lit}. We present the clustercentric radius in units of physical Mpc, where available.}\label{tab:QFE mass lims}
\end{table*}

\subsection{Evolution of the quiescent population in clusters}\label{sec:evolution}

To compare our results with the $z\sim1$ descendents of these protoclusters, we note that the average halo mass growth between $z=2.3$ and $z=1.3$ is about a factor of $5$ \citep{Correa_2015}.  Assuming the total stellar mass grows by the same factor, we predict a total integrated stellar mass, for galaxies with $\log{M_\ast/M_\odot}>9.5$, of log($\rm M_{\ast}/M_{\odot}$) $\sim 12.1^{+0.05}_{-0.06}$ 
for the descendent system at $z \sim 1$. From the total stellar mass to halo mass relation at $z\sim 1$ this corresponds to a halo mass of $\log{\rm M_{200}/M_\odot} \approx 13.7$ \citep{Leauthaud12, vdB14}, corresponding to group-scale haloes. 
We therefore compare with the group sample from \citet{Reeves} at  $1 < z < 1.5$, which have halo masses between $\rm 13.65 \leq log_{10}(M_{200}/M_{\odot}) \leq 13.93$. These groups are selected from the COSMOS \citep[UltraVISTA;][]{McCracken_2012, Muzzin_2013} and SXDF \citep[SPLASH-SXDF;][]{Splash18} photometric surveys, and have some spectroscopic coverage by GOGREEN. These systems have a total integrated stellar mass of $\sim 12.0$ log($\rm M_{\ast}/M_{\odot}$) between $\rm 9.5 < log(M_{\ast}/M_{\odot}) < 11.75$, comparable to the projected mass of our protocluster descendents.

We therefore project our protocluster SMFs to $z=1.3$, assuming they grow by a factor of 5, by 
adding sufficient field galaxies to increase the total stellar mass by this factor.  We assume the field is represented by the  $1 < z < 1.5$ field SMF that we measure in COSMOS.  This ensures that the accreted population evolves identically to the field, with no additional  environmentally-driven evolution. Because of this large mass growth, the projected SMF shape is dominated by that of COSMOS $z\sim 1$ field galaxies.

\begin{figure*}
\centering
\includegraphics[width=\linewidth]{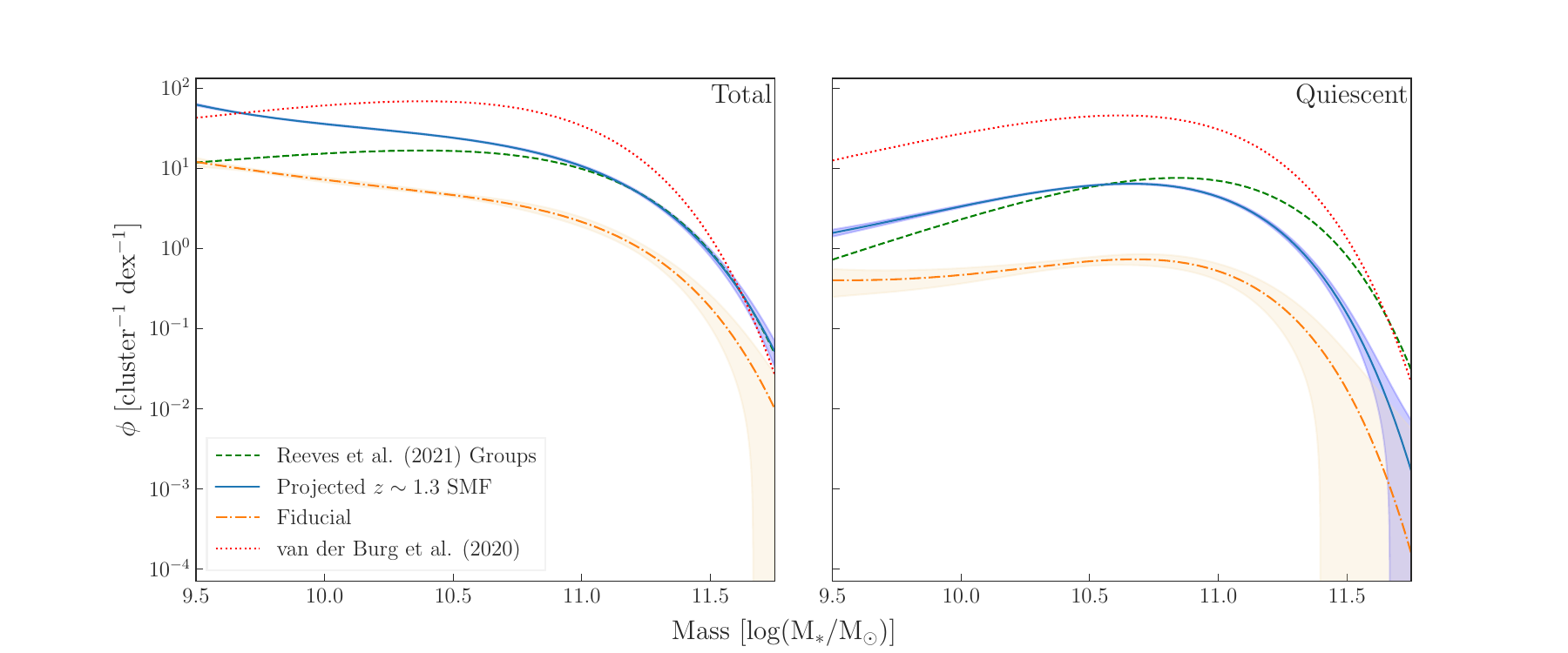}
\caption{The orange curves show our fiducial protocluster SMF, for the total population (left) and quiescent galaxies (right), as presented in Figure~\ref{fig:FSub_SMF_1_02_all}.  We project this to $z=1.3$ by assuming that protoclusters increase in mass by a factor of five through accretion of field galaxies, as described in the text.  This projection is shown as the blue curve.  This is compared with the observed SMF of \citet{Reeves} groups between $1 < z < 1.5$ (green dashed line).
We additionally show the observed SMF of the overall GOGREEN cluster sample between $1 < z < 1.5$ \citep{van_der_Burg_2020} as the red dotted line. 
\textit{Left}: Our projected total SMF, while having a similar total integrated mass as that of the groups, has a different shape, with significantly more low mass galaxies. This implies significant merging and/or disruption of galaxies, as has been found in lower redshift studies \citep{Rudnick+12} and simulations \citep{Bahe+19}. 
\textit{Right}: The shape of the projected quiescent SMF agrees reasonably well with that in the descendent clusters, but with fewer high-mass galaxies. A large number of cluster galaxies at masses above $\rm M_{\ast} > 10^{10.5} M_{\odot}$ must quench star formation in the $2$ Gyr between $z=2.3$ and $z=1.3$. However, for galaxies with masses below $\rm M_{\ast} < 10^{10.5} M_{\odot}$, no additional quenching upon infall is required.
}
\label{fig:projected}
\end{figure*}

The result is shown in Figure~\ref{fig:projected}, compared with both the $1 < z < 1.5$ groups described above, and the more massive GOGREEN-only sample from \citet{van_der_Burg_2020}.
While our projected SMF has a similar normalization to the groups sample, the shapes of the SMFs are 
different, as the observed $z=1$ groups have far fewer low-mass galaxies than the projection.
This may indicate that a significant amount of merging and disruption occurs among cluster members during this time, as expected \citep{Bahe+19}. An alternate explanation would be that clusters don't grow through the unbiased accretion of field galaxies \citep{Lammim+23}.  

When considering just the quiescent population (right panel, Fig.\ref{fig:projected}), the predicted number of high-mass ($M_{\ast}>10^{10.5}~{\rm M}_\odot$) quiescent galaxies in our projection is about five times lower than what is observed in the $z=1$ groups. Additional processes are therefore required to build up massive quiescent galaxies in these groups, with mergers a plausible explanation. However, at lower masses ($M_{\ast}<10^{10.5}~{\rm M}_\odot$), the observed abundance of quiescent galaxies in $z=1$ groups is consistent with, or even larger than, our projections.  This implies that no additional quenching upon infall is required. This is somewhat surprising, as \citet{McNab_2021} found from an analysis of post-starburst galaxies that low mass galaxies in massive $z=1$ clusters have been only quenched recently, upon infall.
This may be evidence for a halo mass dependence on environmental quenching.  The GOGREEN clusters studied in \citet{McNab_2021} are about $\sim 5$ times more massive than the groups in \citet{Reeves}, on average.
Possibly low mass galaxies at this redshift are effectively quenched by environment only when accreted into sufficiently massive structures.

\subsection{Future Work}
Most of the protoclusters included in our sample were originally identified based on an overdensity of star-forming galaxies \citep[e.g.][]{Wang_2016}.  It is possible that this selection is biased against protoclusters with a high fraction of quiescent galaxies.  For example,  QO-1000 only has a  $\sim 1 \sigma$ excess of star-forming galaxies, and would never have been identified as a protocluster by just looking at the star-forming population \citep{Ito_2023}.
Moreover, the COSTCO protoclusters are noted to be mild overdensities at this redshift epoch, and are projected to be relatively low-mass clusters by $z = 0$ \citep{Ata_2022}.  In contrast, we find evidence that the most massive system in our sample, ZFOURGE/ZFIRE, may have a significantly larger quiescent population than the rest of the sample (see Fig. \ref{fig:fsub_ZFORGE}), though the uncertainties are too large to be definitive. Our results are limited by statistics, and much larger samples are therefore needed.
The Euclid deep fields will cover an area $>$ 20 times larger than the COSMOS subset used in this study \citep{Sartoris_2016} and will reach a similar 5$\sigma$ depth (26 mag) as COSMOS for the $Y$, $J$ and $H$ filters \citep{Moneti_2022}. Scaling from our present sample size of 14 protoclusters in a $\sim 2~{\rm deg}^2$ region, the resulting uncertainties on the derived protocluster quiescent fraction can be reduced by a factor of $\sim$5.  Even better results can be expected from the Nancy Grace Roman Space Telescope (NGRST), which will cover an area that is $\sim 40$ times larger than the Euclid Deep Fields, to greater depth.

Additionally, at these high-redshifts where protoclusters lie, better photometric redshift precision is important to reduce line-of-sight uncertainties. The upcoming COSMOS-Web catalogues will help with this, as they are expected to be much deeper, reaching 5$\sigma$ depths of 27.5-28.2 magnitudes in the four NIRCam filters used, spanning $\sim 0.54$ deg$^{2}$ \citep{casey2023cosmosweb}. The Hyperion structure lies in this field, and its fainter members are expected to be mapped for the first time \citep{casey2023cosmosweb}.  Ultimately, however, the largest gains will be with spectroscopy.  A deep slitless prism survey with NGRST covering just 10 square degrees could identify  $\sim 15,000$ protocluster galaxies \citep{Roman-wp}.  

\section{Conclusions}

We measure the SMFs for the total, star-forming and quiescent galaxy populations in protoclusters between $2 < z  < 2.5$ in the COSMOS field.  These are compared with a comparably selected field sample and are estimated to be complete down to a mass limit of log$(M_{\rm lim}/M_{\odot})=9.5$. We use these SMFs to measure the efficiency of environmental quenching in protocluster regions as opposed to the field. Our main findings are:

\begin{itemize}[wide, labelwidth=!,itemindent=!,labelindent=0pt, leftmargin=0em]
\item On a scale of $1$ Mpc we find a significant overdensity of galaxies in fields centred on the protocluster sample.  The shape of the protocluster total SMF, and that of the dominant, star-forming population, is consistent with that of the field (Fig. \ref{fig:SMF_1_02_all}). 
\item The shape of the protocluster quiescent SMF is different from the field. 
It is flatter than the field at low masses, with a relative excess of galaxies $M_{\ast}<10^{10} M_{\odot}$ (Fig. \ref{fig:FSub_SMF_1_02_all}). This difference is only significant at a $\sim2\sigma$ level, however (Fig~\ref{fig:phi vs alpha}).
\item The fraction of quenched galaxies in our fiducial protocluster selection is indistinguishable from the field above $M_{\ast}>10^{10}M_{\odot}$.  However, there is a small but significant excess ($0.08^{+0.03}_{-0.02}$) at lower masses (Fig. \ref{fig:Fsub_QF}). 
\item  We compare the protoclusters with a sample of groups at $1<z<1.5$ from \citet{Reeves}.
The total stellar mass of those groups within $1$ Mpc is about a factor $\sim 5$ larger than in the protoclusters. This is similar to the expected mass growth over this time \citep{Correa_2015}. 
We calculate a projected descendent SMF by assuming the protoclusters grow via accretion of field galaxies to the mass of the group sample, with no additional quenching.  The resulting shape of this projected SMF is significantly different from that of the lower redshift sample, with an excess of galaxies with $M_{\ast} \lesssim 10^{11} M_{\odot}$  (Fig. \ref{fig:projected}, left panel). This may indicate that significant merging and/or disruption of galaxies takes place between $z=2.3$ and $z=1.3$. 
\item To match the observed quiescent SMF in the group sample, the number of massive quiescent galaxies must increase by about a factor $\sim 5$ beyond what is predicted from the accretion of field galaxies, between $1.3 \lesssim z \lesssim 2.3$ (Fig. \ref{fig:projected}, right panel). However, at low masses ($\rm M_{\ast} < 10^{10.5}M_{\odot}$), no additional quenching upon accretion is necessary, and in fact we project even more low-mass galaxies than are observed in \citep{Reeves}. This is surprising, as \citet{McNab_2021} shows that in the GOGREEN clusters, low-mass galaxies are expected to have quenched more recently than high-mass ones. This can plausibly be explained by the much larger halo mass of the GOGREEN  clusters having more of an environmental effect on low mass galaxies.

\end{itemize}

We conclude that the SMF of galaxies within $1$ Mpc of these protoclusters is similar to that of the field, with a small fraction of quiescent galaxies.  There is some evidence for a small excess of low-mass ($\log{M/M_\odot}<10$) quiescent galaxies relative to the field, though this is of modest significance and may also be impacted by incompleteness.   
In any case,
these are small in number, and most of the quiescent galaxies that dominate rich clusters at $1<z<1.5$ must therefore have been quenched later, presumably upon accretion \citep[though see][for an alternative explanation]{Lammim+23}.  The lack of massive, primordially-quenched galaxies is a surprise given the results of some other studies \citep[e.g.][]{McConachie_2022}.  As we rely on photometric redshifts with statistical background subtraction, the uncertainties for this small sample of 14 protoclusters are large, especially for $M_{\ast}>10^{11.0}M_{\odot}$.
Future studies based on larger samples (e.g. from the Euclid deep fields) and with more precise and accurate redshifts (e.g. from COSMOS-Web) should significantly improve upon these results.


\section*{Acknowledgements}
We thank the referees for their useful comments.
This work was developed as part of an ISSI workshop on protoclusters, held in Bern, Switzerland in 2022.  We are grateful for the support of ISSI and the use of their facilities.  MB acknowledges support from the NSERC Discovery Grant program.
Based on observations collected at the European Southern Observatory under ESO programme ID 179.A-2005 and on data products produced by CALET and the Cambridge Astronomy Survey Unit on behalf of the UltraVISTA consortium.  
MCC acknowledges support from the National Science Foundation through grant AST-1815475.
RD gratefully acknowledges support by the ANID BASAL project FB210003.  GHR acknowledges support from NSF-AST grants 1517815 and 2206473 as well as HST grants AR-14310.001-A GO-15294.012-A and NASA ADAP grant 80NSSC19K0592.  BV acknowledges support from the INAF Mini Grant 2022 "Tracing filaments through cosmic time" (PI Vulcani).
GC acknowledges the support from the grant ASI
n.2018-23-HH.0.
GW gratefully acknowledges support from NSF grant AST-2205189 and from HST program number GO-16300. Support for program number GO-16300 was provided by NASA through grants from the Space Telescope Science Institute, which is operated by the Association of Universities for Research in Astronomy, Incorporated, under NASA contract NAS5-26555.

\section*{Data Availability}

All data used in this paper are obtained from the published catalogue of \citet{Weaver_2022}. 

\textit{Software}: Astropy \citep{astropy:2022}, Corner \citep{corner}, Emcee \citep{Foreman_Mackey_2013}, Matplotlib \citep{matplotlib}, Numpy \citep{numpy}, SciPy \citep{Scipy}, Seaborn \citep{Waskom2021}.



\bibliographystyle{mnras}
\bibliography{main} 
\input{appendices.tex}
\bsp	
\label{lastpage}
\end{document}

%% file: appendices.tex

\appendix
\section{Field Considerations}\label{Field considerations}
\subsection{Redshift distribution of the Field} \label{Field Choice}
In this work, we define our `field' to be all objects that match our cut selection (see Section \ref{Sample Selection}) between $2 < z < 2.5$. This differs from the range in which our candidate protocluster members are selected, $1.8 < z < 2.7$.  The broader range is necessary to accommodate photometric redshift uncertainties (see Section \ref{photoz}).  The normalization of the field SMF is somewhat sensitive to the redshift range, and we choose the narrower $2<z<2.5$ to better match the redshift distribution of the protocluster members.  In particular, there are overdense structures at $1.8<z<2$ that significantly perturb the field SMF when that range is included.

\begin{figure}
\centering
\includegraphics[width=1\linewidth]{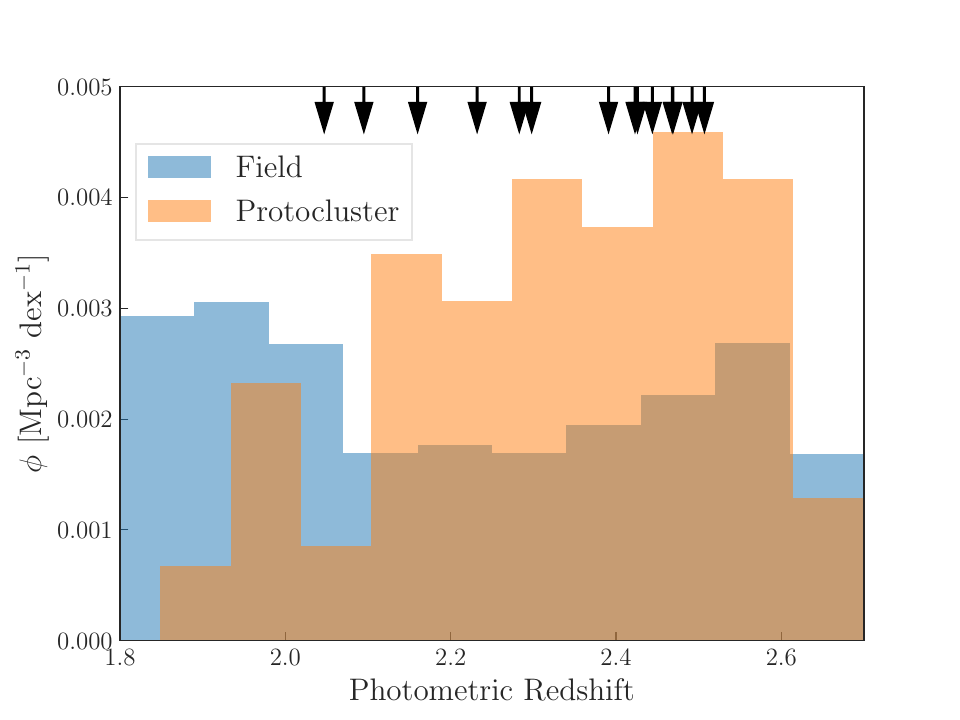}
\caption{The distribution of field and protocluster (fiducial selection, Table \ref{tab:selection_quantities}) photometric redshifts. Arrows indicate individual protocluster candidate redshifts (Table \ref{tab:Protocluster_candidates}). As can be seen, the majority of protocluster galaxies lie in the $2 < z < 2.5$ redshift range, as expected. Selecting a field sample in this same range helps to ensure a similar redshift distribution as the clusters, by avoiding the overdense structures between $1.8 < z < 2$.}
\label{fig:photoz dist}
\end{figure}
\clearpage

\subsection{Field Comparison}\label{app-field}
\begin{figure*}
\centering
\includegraphics[width=1\linewidth]{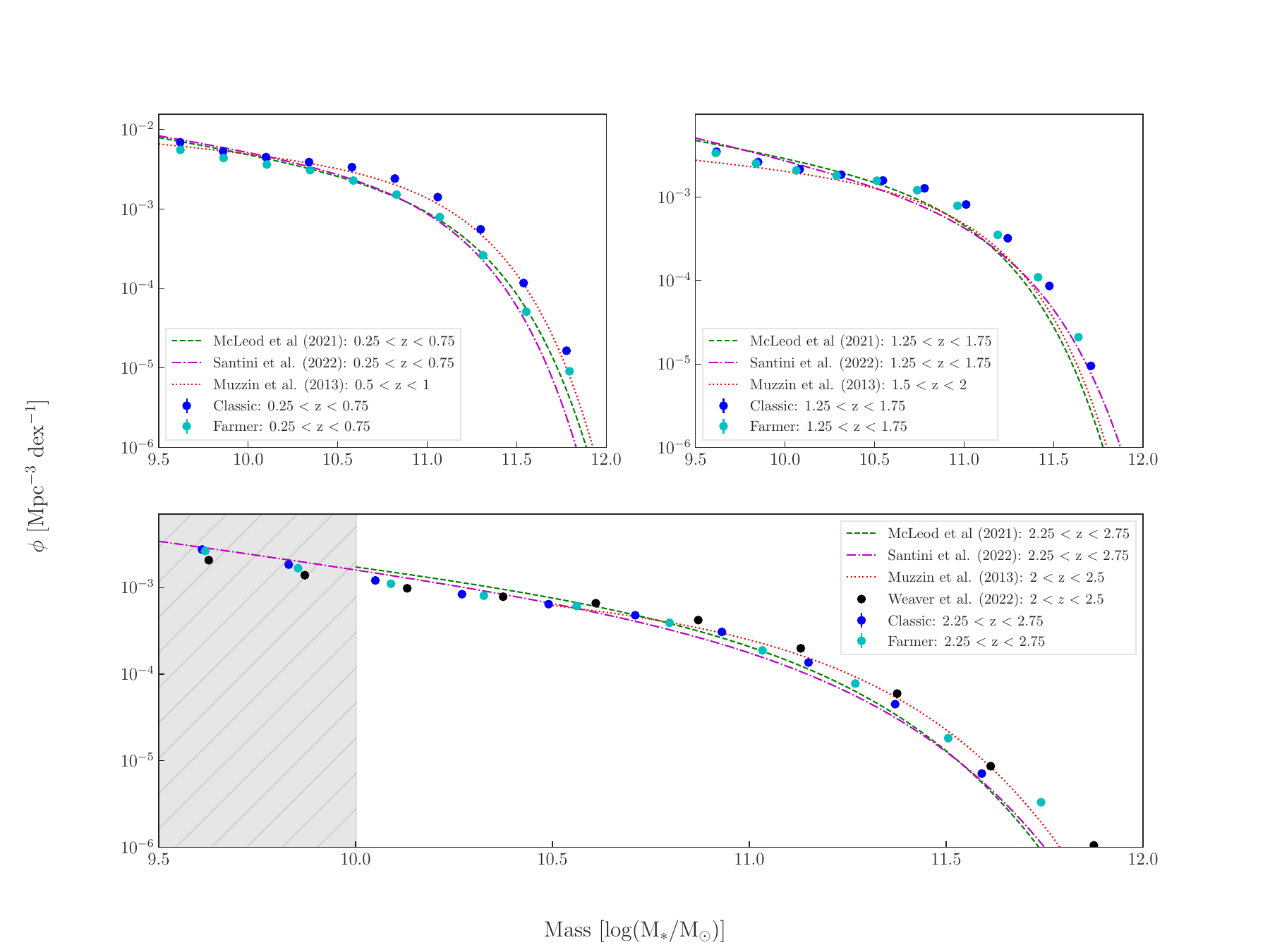}
\caption{We show the total SMF of our field sample in three redshift bins, chosen to correspond to those of \citet{McLeod_2021} and \citet{Santini_2022}.  There is good agreement for $M_{\ast}< 10^{11} M_{\odot}$.}
\label{fig:lit_comp}
\end{figure*}

In Figure~\ref{fig:rel_lit_comp} we compared the total and quiescent SMFs in our field sample to that measured in previous studies.   As noted in the text, there are some differences in how our sample is constructed relative to those comparison studies. Here we show the extent to which these differences affect the SMF measurements.  First, in Figure~\ref{fig:lit_comp} we show the total SMF in three different redshift bins, $0.25<z<0.75$, $1.25<z<1.75$ and $2.25<z<2.75$.  The latter bin is different from our default field sample ($2<z<2.5$), chosen here to correspond to the binning of \citet{McLeod_2021} and \citet{Santini_2022}.   In general, the total SMF agrees well with both those works, as well as that of \citet{Muzzin_2013}, in all three redshift ranges, for $M_{\ast}< 10^{11} M_{\odot}$.  We also show the effect of using our default photometric catalogue, \verb|Farmer|, compared with the \verb|Classic|. The difference is largely negligible, especially at the $z\approx 2$ epoch that is central to this work.

\begin{figure*}
\centering
\includegraphics[width=1\linewidth]{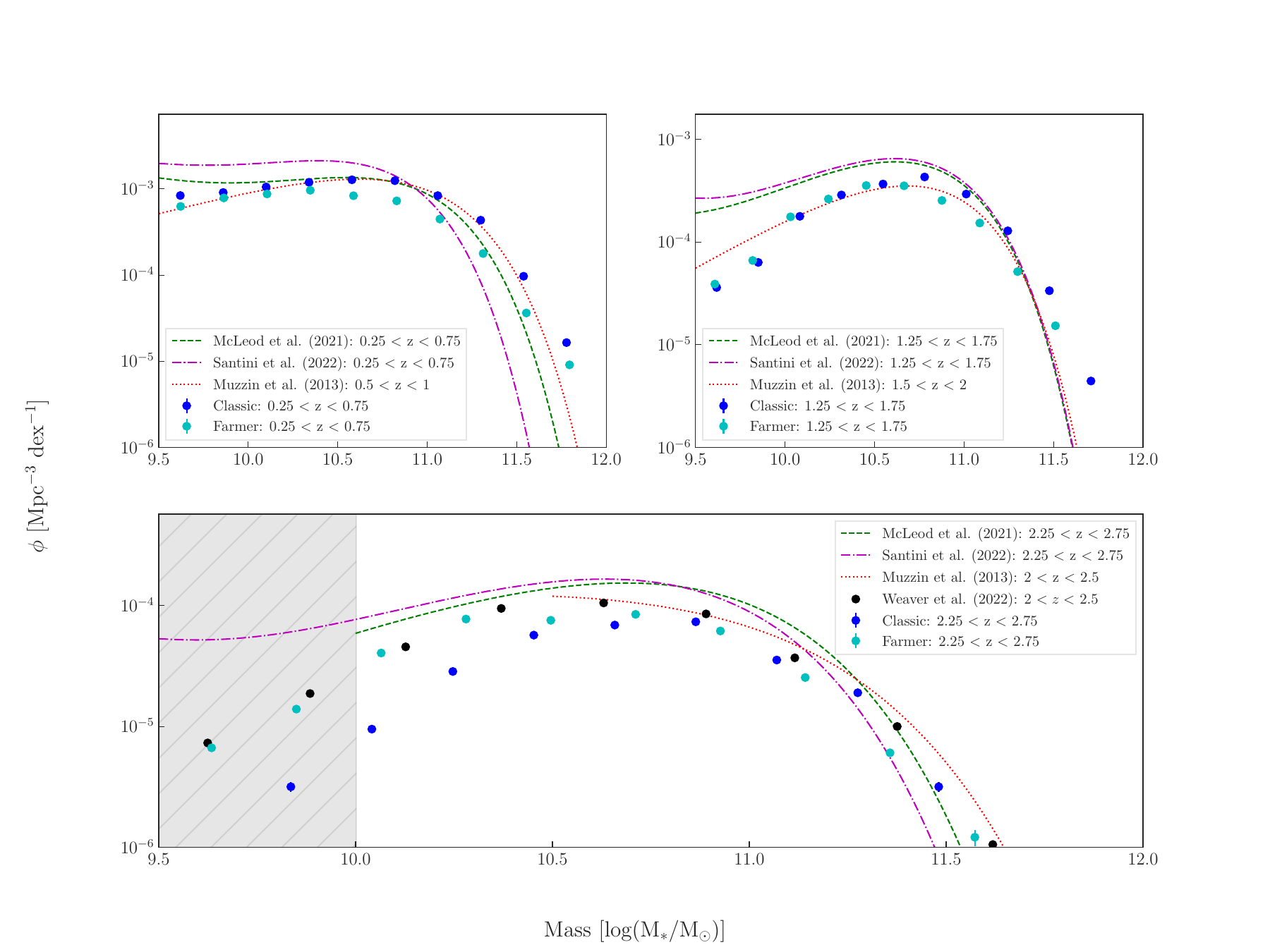}
\caption{Similar to Figure~\ref{fig:lit_comp}, but for the quiescent population.   As with our fiducial sample in Figure~\ref{fig:rel_lit_comp2}, the normalization of our observed SMF is lower than that of \citet{McLeod_2021} and \citet{Santini_2022}, even when using the same redshift bin of $2.25<z<2.75$.  This difference also persists at lower redshift.}  
\label{fig:lit_comp2}
\end{figure*}

Figure~\ref{fig:lit_comp2} is similar, but for just the quiescent population.  The lower normalization that we observe relative to \citet{McLeod_2021} and \citet{Santini_2022} persists even when we use the same redshift interval $2.25<z<2.75$, and also extends to the lower redshift interval $1.25<z<1.75$.  We also observe that the choice of catalogue (the default, \verb|Farmer|, compared with the \verb|Classic|) has a significant impact on the quiescent SMF at low stellar masses, as discussed by \citetalias{Weaver_2022} and \citetalias{Weaver_SMF}.  

\begin{figure*}
\centering
\includegraphics[width=1\linewidth]{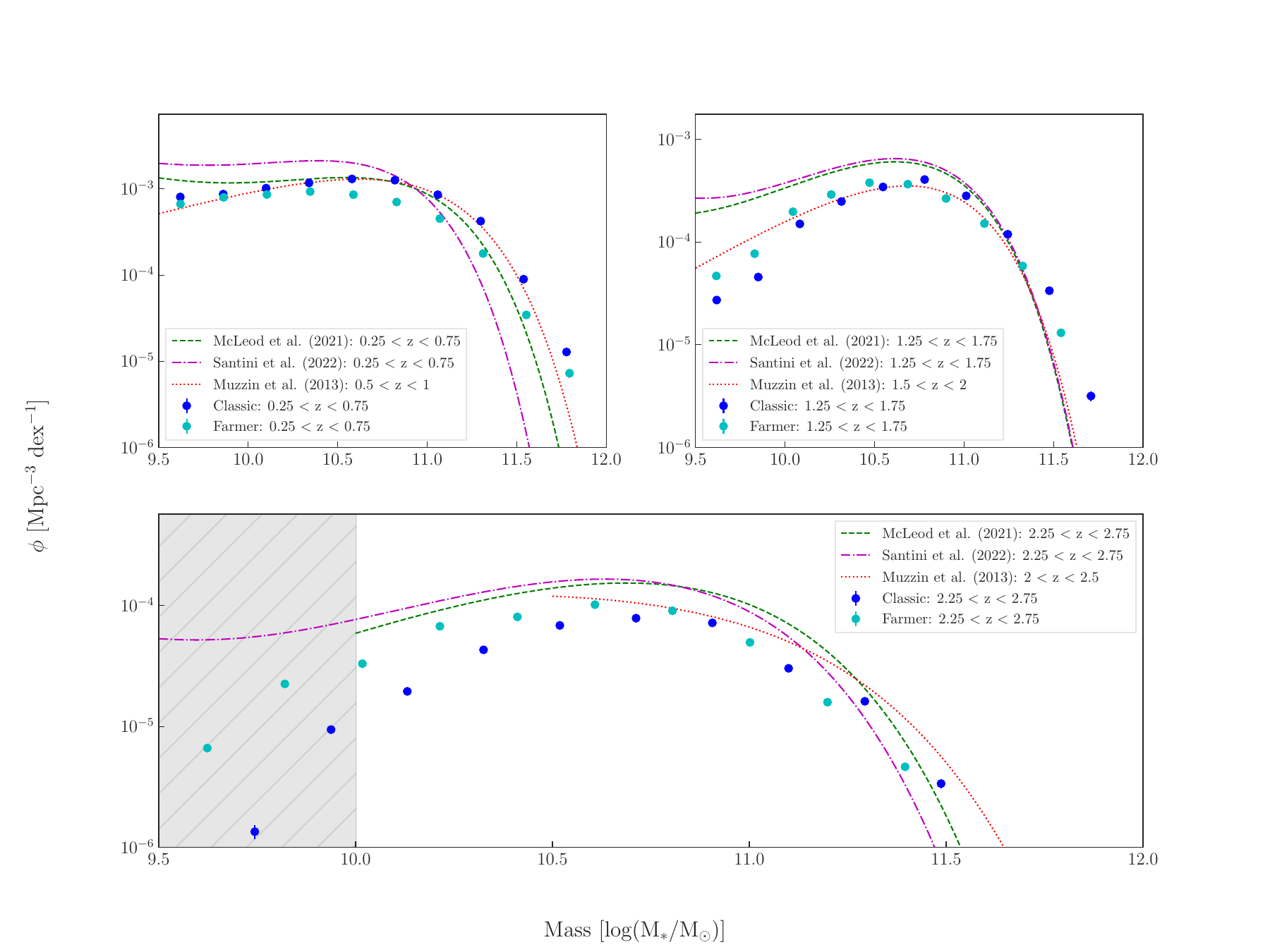}
\caption{As Figure~\ref{fig:lit_comp2}, but where quiescent galaxies in our sample are defined from their UVJ colours, rather than the default NUVrJ.  This does not remove the discrepancy with \citet{McLeod_2021} or \citet{Santini_2022}, who also use a UVJ colour classification.}
\label{fig:UVJ_lit_comp}
\end{figure*}

Finally, we consider the impact of selecting quiescent galaxies from a UVJ colour selection, rather than our default NUVrJ.  The result is shown in Figure~\ref{fig:UVJ_lit_comp}, for the same redshift bins as the previous two figures.  Again, the different choice in definition does not remove the discrepancy with \citet{McLeod_2021} and \citet{Santini_2022}. 

\clearpage

\section{Stellar mass functions in different volumes} \label{Backgroun Subs}
\begin{figure*}
\centering
\includegraphics[width=.95\linewidth, trim = 150 0 250 0]{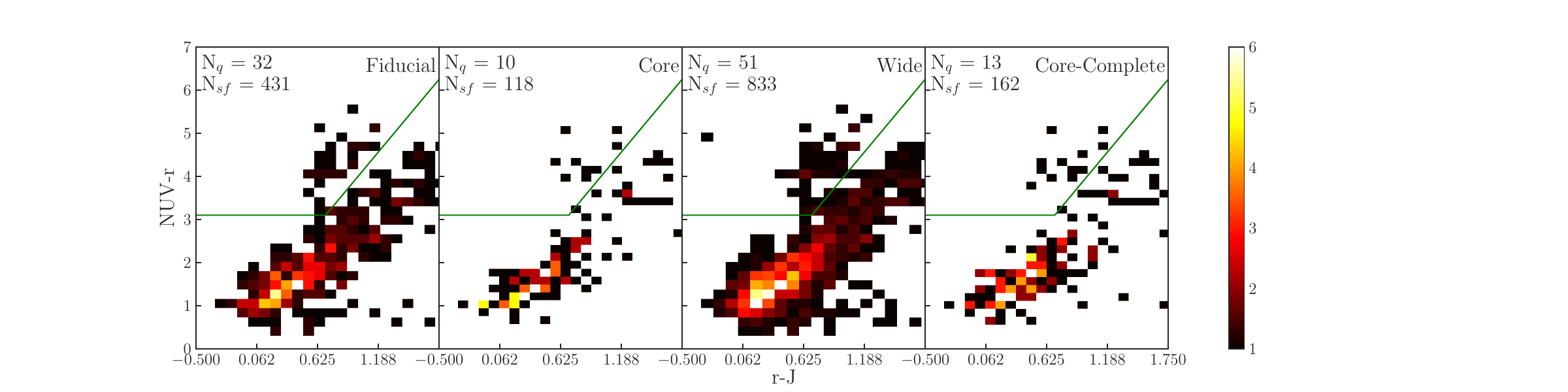}
\caption{The (NUV-$r$) vs. ($r$-$J$) colour distribution is shown for each volume selection (Table~\ref{tab:selection_quantities}).  The division between quiescent and star-forming galaxies is shown as the solid line.   We observe a distinct quiescent population in each selection volume.}.

\label{fig:mult_NUVrJ}
\end{figure*}

In this section we show how the intrinsic protocluster SMFs depend on the different protocluster volume selections tabulated in Table~\ref{tab:selection_quantities}. These can be compared with our fiducial results, in Figure~\ref{fig:FSub_SMF_1_02_all}.  The NUVrJ colour distributions of each sample are shown in  Figure~\ref{fig:mult_NUVrJ}.
The morphology of the colour distribution is similar for all samples, with the primary difference being one of sample size. 

\begin{figure*}
\centering
\includegraphics[width=1\linewidth]{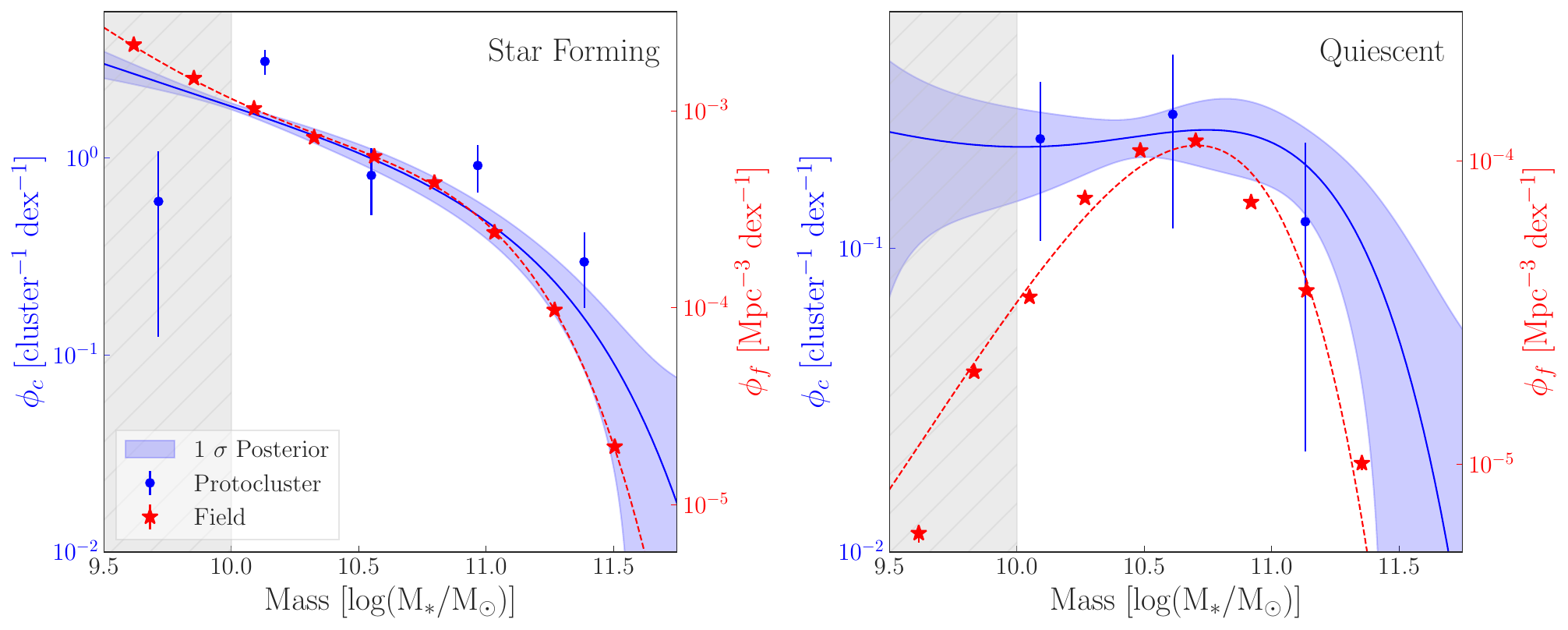}
\caption{We show the intrinsic protocluster SMFs for our Core selection ($dR = 0.5$ Mpc, $dz = 0.2$), to be compared with our fiducial results in Figure~\ref{fig:FSub_SMF_1_02_all}.  In this sample, the excess of low-mass quiescent galaxies is even more pronounced, with an SMF that increases steeply toward lower masses.  There is also an excess of massive, quiescent galaxies, and a deficit of low-mass, star-forming galaxies.}
\label{fig:FSub_SMF_05_02_all}
\end{figure*}

First, in Figure~\ref{fig:FSub_SMF_05_02_all} we show the Core sample, $dR, dz = 0.5$ Mpc, $0.2$.  As with the fiducial sample, we observe a significant excess of low-mass protocluster galaxies, with an SMF that rises even more steeply towards lower masses.  In addition, there is a bump at $M_{\ast} \sim 10^{11.25} M_{\odot}$, corresponding to an excess of very massive galaxies that is not seen in the wider selection.  Also different from the fiducial sample is the drop in the number of star-forming (and, hence, total) galaxies at the lowest stellar masses.

\begin{figure*}
\centering
\includegraphics[width=1\linewidth]{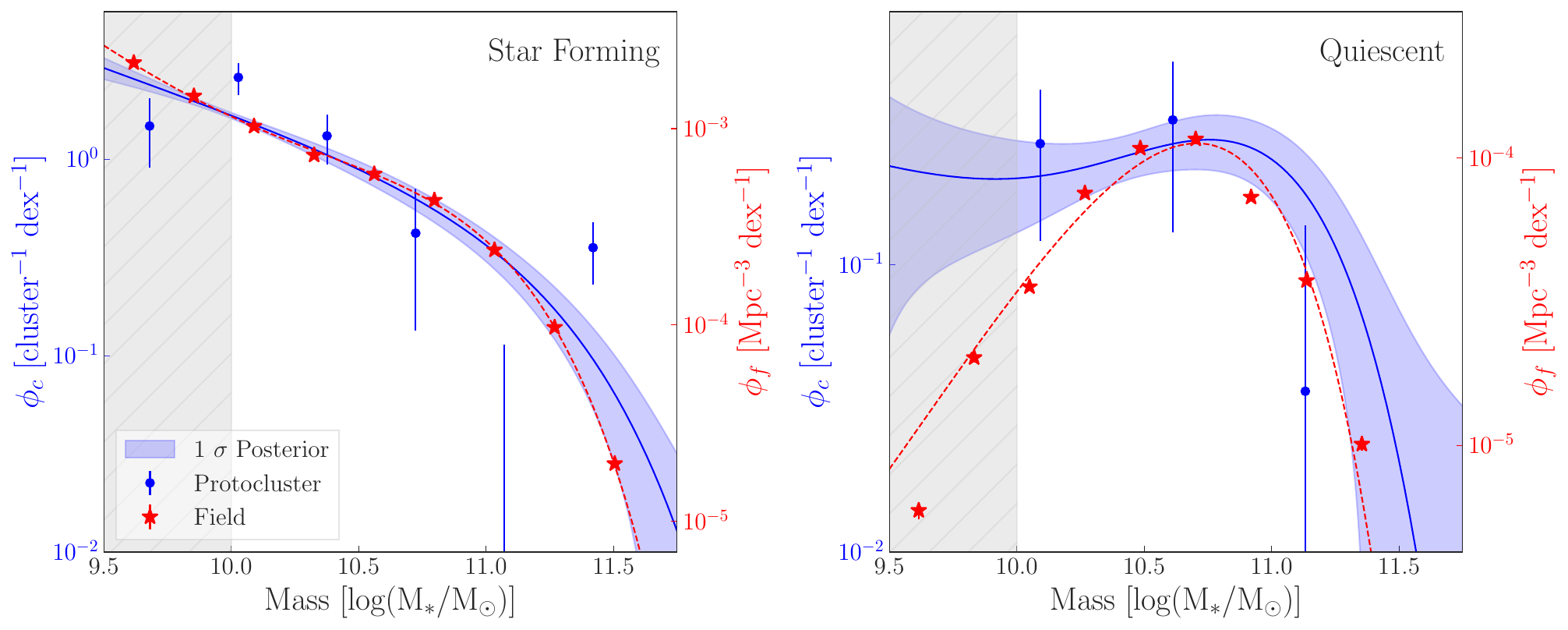}
\caption{As Figure~\ref{fig:FSub_SMF_1_02_all}, but for the Core-complete sample ($dR = 0.5$ Mpc, $dz = 0.3$).  Results are very similar to the Core sample shown in Figure~\ref{fig:FSub_SMF_05_02_all}.  Uncertainties on the quiescent SMF are larger because the larger $dz$ results in greater field contribution within the volume.  }
\label{fig:FSub_SMF_05_03_all}
\end{figure*}

Next in Figure~\ref{fig:FSub_SMF_05_03_all} we consider the Core-complete selection ($dR, dz = 0.5, 0.3$). This is similar to the Core sample just discussed, but with a higher completeness due to the larger $dz$ range, chosen to include $\sim 95\%$ of all quiescent galaxies in the cluster (See section \ref{photoz}).  The results are generally indistinguishable from Figure~\ref{fig:FSub_SMF_05_02_all}, though the uncertainties on the quiescent SMF are  larger due to the increased field contamination.  This demonstrates that the narrower $dz=0.2$ selection used in our fiducial sample does not significantly bias the results against quiescent galaxies.

\begin{figure*}
\centering
\includegraphics[width=1\linewidth]{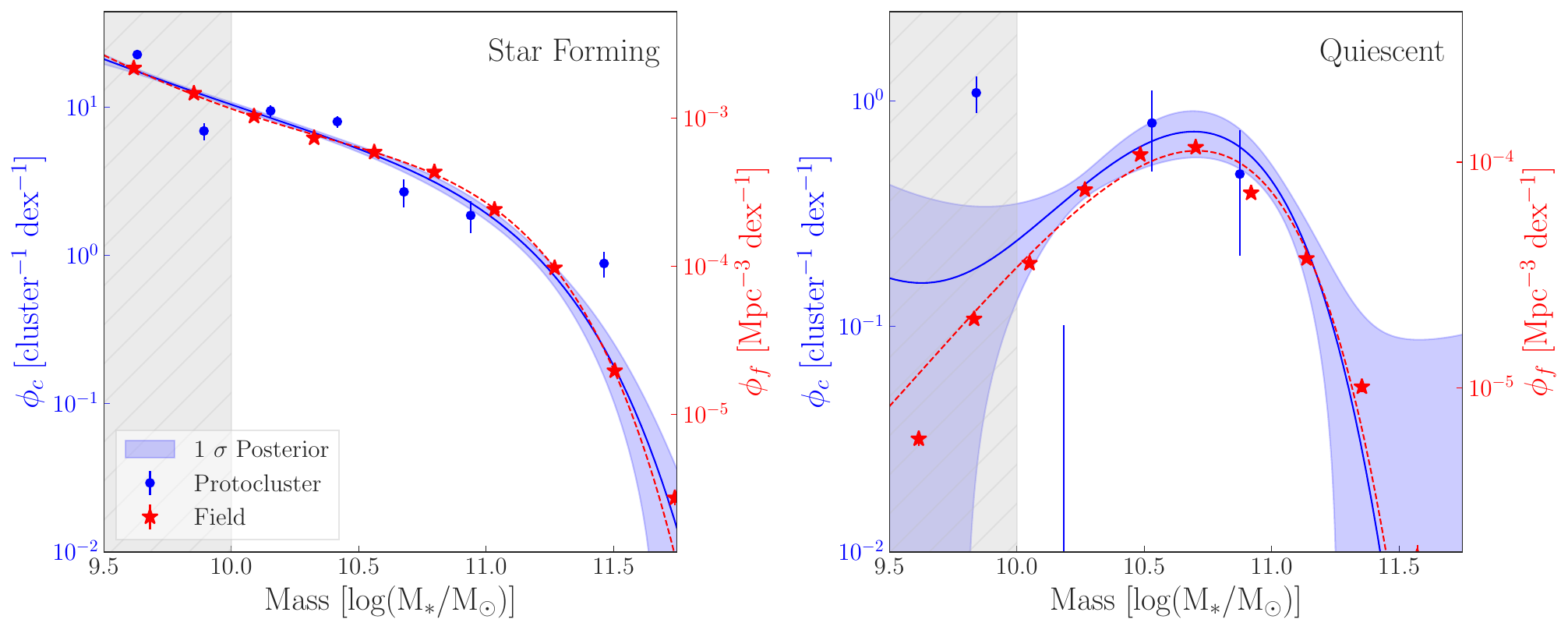}
\caption{As Figure~\ref{fig:FSub_SMF_1_02_all}, but for the Wide selection ($dR = 1.5$ Mpc, $dz = 0.2$).  The SMF shapes are generally consistent with the field, though the flatter shape of the quiescent SMF in protoclusters is still present.}
\label{fig:FSub_SMF_15_02_all}
\end{figure*}

Finally, in Figure~\ref{fig:FSub_SMF_15_02_all}, we show the Wide selection of $dR=1.5$ Mpc and $dz = 0.2$.  For this volume, the SMFs are in general much more similar in shape to that of the field.  An excess of low-mass quiescent galaxies is still apparent, though it is not statistically significant.

We also look at the intrinsic protocluster SMF for galaxies with selection parameters $dR = 1 Mpc, dz =0.2$ around just the most massive protocluster in our sample, ZFOURGE/ZFIRE. We note that there are no quiescent galaxies above $10^{10}M_{\odot}$. However, there is a large low-mass excess in this protocluster; all six quiescent galaxies in this selection have masses below $M_{\ast} < 10^{10.6} M_{\odot}$.  Notably, this number of quiescent galaxies is about an order of magnitude larger than the average per cluster when considering the full sample.

\begin{figure*}
\centering
\includegraphics[width=1\linewidth]{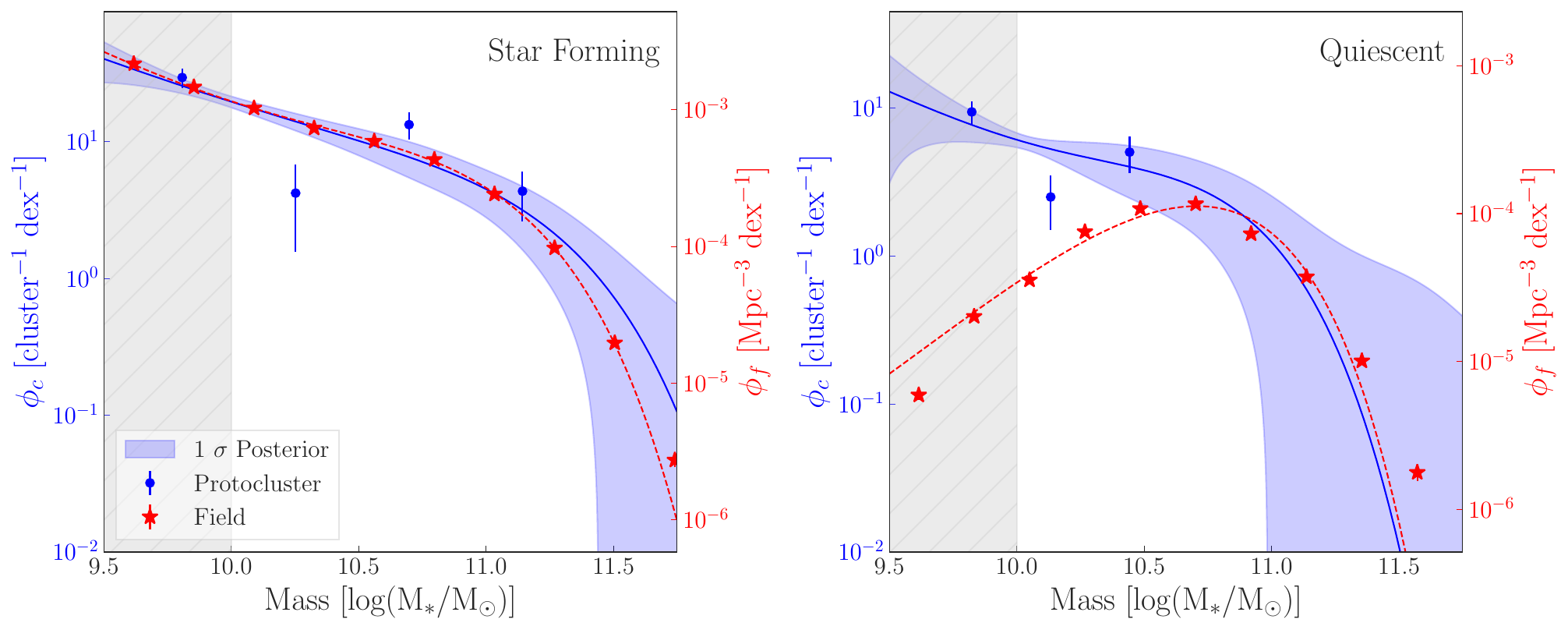}
\caption{As Figure~\ref{fig:FSub_SMF_1_02_all}, but for just galaxies with selection parameters $dR = 1$ Mpc, $dz = 0.2$ around just the most massive protocluster in our sample, ZFOURGE/ZFIRE. The number of low-mass quiescent galaxies here is about a factor ten larger than the average for our full sample.}
\label{fig:fsub_ZFORGE}
\end{figure*}

We present the fit parameters for the intrinsic protocluster SMFs in each selection, as well as the fit parameters for the field in Table~\ref{tab:vals}.
\begin{table*}
\begin{tabular}{@{}llllllll@{}}
\toprule
\textbf{Selection} & \textbf{Alias} & \textbf{Population} & \textbf{$\mathcal{M}$*} & \textbf{log$\phi^{\ast}_{1}$} & \textbf{$\alpha_1$} & \textbf{log$\phi^{\ast}_{2}$} & \textbf{$\alpha_{2}$} \\ \midrule
      &            & Total               & 10.91$^{+0.19}_{-0.16}$        & 0.34$^{+0.20}_{-0.24}$           & -0.72$^{+0.41}_{-0.63}$            & -0.34$^{+0.26}_{-0.28}$          & -1.64$^{+0.36}_{-0.27}$            \\
A  & Fiducial      & Quiescent           & 10.66$^{+0.21}_{-0.18}$        & -0.15$^{+0.11}_{-0.14}$          & 0.16$^{+0.61}_{-0.52}$             & -1.11$^{+0.32}_{-0.75}$          & -1.21$^{+0.15}_{-0.43}$            \\
       &           & Star Forming        & 10.95$^{+0.19}_{-0.16}$        & 0.28$^{+0.17}_{-0.28}$           & -0.93$^{+0.43}_{-0.38}$            & -0.74$^{+0.47}_{-1.07}$          & -1.88$^{+0.26}_{-0.75}$            \\ \midrule
            &      & Total               & 11.20$^{+0.31}_{-0.21}$        & -0.45$^{+0.27}_{-0.31}$          & -1.12$^{+0.54}_{-0.30}$            & -1.10$^{+0.34}_{-0.51}$          & -1.53$^{+0.26}_{-0.18}$            \\
B  & Core          & Quiescent           & 10.95$^{+0.41}_{-0.71}$        & -0.66$^{+0.17}_{-0.28}$          & -0.12$^{+0.95}_{-0.71}$            & -1.54$^{+0.30}_{-0.57}$          & -1.27$^{+0.18}_{-0.30}$            \\
           &       & Star Forming        & 11.22$^{+0.37}_{-0.17}$        & -0.58$^{+0.19}_{-0.30}$          & -1.24$^{+0.51}_{-0.21}$            & -1.29$^{+0.34}_{-0.45}$          & -1.59$^{+0.38}_{-0.19}$            \\ \midrule
          &        & Total               & 10.91$^{+0.21}_{-0.11}$        &  0.47$^{+0.15}_{-0.24}$          & -0.91$^{+0.30}_{-0.46}$            & -0.51$^{+0.42}_{-0.58}$          & -1.90$^{+0.32}_{-0.33}$            \\
C  & Wide          & Quiescent           & 10.49$^{+1.32}_{-0.14}$        & -0.17$^{+0.05}_{-0.39}$          &  0.60$^{+0.59}_{-0.66}$            & -2.05$^{+0.82}_{-1.14}$          & -1.81$^{+0.65}_{-0.95}$            \\
         &         & Star Forming        & 11.00$^{+0.17}_{-0.14}$        & 0.24$^{+0.20}_{-0.23}$           & -1.07$^{+0.35}_{-0.40}$            & -0.43$^{+0.32}_{-0.56}$          & -1.79$^{+0.13}_{-0.28}$            \\ \midrule
        &          & Total               & 11.19$^{+0.24}_{-0.19}$        & -0.54$^{+0.23}_{-0.27}$          & -1.27$^{+0.41}_{-0.19}$            & -1.18$^{+0.33}_{-0.48}$          & -1.50$^{+0.28}_{-0.19}$            \\
D  & Core Complete & Quiescent           & 10.77$^{+0.43}_{-0.25}$        & -0.60$^{+0.12}_{-0.22}$          & 0.17$^{+0.82}_{-0.72}$             & -1.48$^{+0.29}_{-0.79}$          & -1.27$^{+0.18}_{-0.37}$            \\
       &           & Star Forming        & 11.25$^{+0.31}_{-0.23}$        & -0.75$^{+0.27}_{-0.34}$          & -1.39$^{+0.43}_{-0.19}$            & -1.39$^{+0.36}_{-0.51}$          & -1.57$^{+0.37}_{-0.20}$            \\ \midrule
      &        & Total               & 10.87$^{+0.04}_{-0.03}$        & -3.31$^{+0.02}_{-0.03}$          & -0.74$^{+0.12}_{-0.18}$            & -4.44$^{+0.21}_{-0.56}$          & -2.05$^{+0.12}_{-0.34}$            \\
Field &        & Quiescent           & 10.54$^{+0.03}_{-0.03}$        & -3.92$^{+0.002}_{-0.004}$          & 0.46$^{+0.09}_{-0.09}$             & -6.88$^{+0.53}_{-0.74}$          & -1.30$^{+0.23}_{-0.48}$            \\
       &       & Star Forming        & 10.90$^{+0.05}_{-0.05}$        & -3.45$^{+0.03}_{-0.04}$          & -0.80$^{+0.26}_{-0.17}$            & -4.37$^{+0.31}_{-0.37}$          & -1.98$^{+0.16}_{-0.20}$           \\ \bottomrule
\end{tabular}
\caption{Summary of best-fit parameters for the double Schechter functions fit to each selection and population. The field is defined as everything between $2 < z < 2.5$, and the fit to the intrinsic protocluster SMF $\phi_{c}$ are as described in Section~\ref{sec:methods}.   $\mathcal{M}^{*}$ is in units of log($M_{\ast}/M_{\odot}$), and $\alpha_1$ and $\alpha_2$ are unitless. 
$\phi^{\ast}_1$ and $\phi^{\ast}_2$ are in units of dex$^{-1}$cluster$^{-1}$, except for the field, where it is presented in units of dex$^{-1}$Mpc$^{-3}$.}
\label{tab:vals}
\end{table*}
